\documentclass[twocolumn,aps,prb,10pt]{revtex4-1}

\usepackage{color}
\usepackage{bm}
\usepackage{graphics}
\usepackage{graphicx}
\usepackage{amsthm,amsmath,amssymb,amsfonts}
\usepackage{bbm} 
\usepackage[mathscr]{euscript} 
\usepackage[export]{adjustbox}

\usepackage{animate} 

\usepackage{marvosym} 

\usepackage{array,makecell,multirow,rotating} 


\newcommand{\e}{\mathrm{e}}
\newcommand{\MM}{{\bf M}}
\newcommand{\nn}{{\bf n}}

\newcommand{\st}{\mathrm{t}}
\newcommand{\GG}{\mathrm{G}}

\newcommand{\JJ}{{\bf J}}

\definecolor{darkred}{RGB}{139,0,0}
\definecolor{crimson}{RGB}{220,20,60}
\definecolor{lightblue}{RGB}{173,216,230}
\definecolor{pink}{RGB}{255,192,203}

\begin{document}

\title{\bf Holographic Dynamics from Multiscale Entanglement Renormalization Ansatz}
\date{\today}
\author{Victor Chua, Vasilios Passias, Apoorv Tiwari, Shinsei Ryu}
\affiliation{Department of Physics, and Institute for Condensed Matter Theory, University of Illinois at Urbana-Champaign, Urbana, IL 61801}

\begin{abstract}
The Multiscale Entanglement Renormalization Ansatz (MERA) is a tensor network based variational ansatz that is capable of capturing many of the key physical properties of strongly correlated ground states such as criticality and topological order. MERA also shares many deep relationships with the AdS/CFT (gauge-gravity) correspondence by realizing a UV complete holographic duality within the tensor networks framework. Motivated by this, we have re-purposed the MERA tensor network as an analysis tool to study the real-time evolution of  the 1D transverse Ising model in its low energy excited state sector. We performed this analysis by allowing the ancilla qubits of the MERA tensor network to acquire quantum fluctuations, which yields a unitary transform between the physical (boundary) and ancilla qubit (bulk) Hilbert spaces. This then defines a reversible quantum circuit which is used as a `holographic transform' to study excited states and their real-time dynamics from the point of the bulk ancillae. In the gapped paramagnetic phase of the transverse field Ising model, we demonstrate the holographic duality between excited states induced by single spin-flips (Ising `magnons') acting on the ground state and single ancilla qubit spin-flips. The single ancillae qubit excitation is shown to be stable in the bulk under real-time evolution and hence defines a stable holographic quasiparticle which we have named the `hologron'. Their bulk 2D Hamiltonian, energy spectrum and dynamics within the MERA network are studied numerically. The `dictionary' between the bulk and boundary is determined and realizes many features of the holographic correspondence in a non-CFT limit of the boundary theory. As an added spin-off, this dictionary together with the extension to multi-hologron sectors gives us a systematic way to construct quantitatively accurate low energy effective Hamiltonians. 
\end{abstract}

\maketitle

\section{Introduction and Motivation}

The Multiscale Renormalization Ansatz (MERA) was initially conceived by Vidal\cite{vidal2007entanglement,vidal2009entanglement} dually as an entanglement renormalization scheme and a variational ansatz. Nevertheless it is most usually regarded in practice as a tensor network variational ansatz for complex interacting ground states. Quite remarkably it has an interpretation as a gravitational dual to a conformal field theory (CFT) under the Anti de-Sitter / Conformal Field Theory (AdS/CFT) holographic conjecture.  \cite{ryu2006aspects,ryu2006holographic,swingle2012entanglement,molina2015information,van2009comments,van2010building,swingle2010mutual,molina2011holographic,molina2011connecting,balasubramanian2012momentum,matsueda2011scaling,ishihara2013refined,czech2015tensor} In this interpretation, which has found popular use in the area of (computational) quantum simulation,\cite{evenbly2009algorithms,corboz2010simulation,evenbly2014class,Time_evo_MERA_Matteo} where a quantum circuit defined by the tensors of a MERA acts to manufacture a variational wavefunction by introducing entanglement through its action on a simpler reference wavefunction in a scale by scale manner. The reference wavefunction is often taken to be a product state \footnote{In some generalizations to topologically ordered states, a small superposition of product states is used instead.} and the MERA effectively dresses it with complexity to produce a viable variational ground state. Much developmental work has then been invested into algorithms that optimize the fidelity of variational ground state to the actual ground state; whereby the collection of tensors in a MERA are considered as an efficient representation of the ground state. In fact all tensor network based methods \cite{Or�s_MPS&PEPS_Intro,DMRG_Intro} such as DMRG, MPS and PEPS share this common desired goal of achieving efficient representations of wavefunctions --- more generally mixed density matrices -- through tensors. 

The major conceptual advancement of the MERA scheme is to envisage that it defines a quantum circuit that implements a Renormalization Group (RG) transformation which is organized like a space-time. This seems natural given that the quantum gates (tensors) in the network act on local degrees of freedom defined on the network. By dividing the network into RG iteration steps, we can then define a global foliation or time-slicing. This organization of tensors into a discrete space-time leads to emergent bulk degrees of freedom organized by a causal structure, where `earlier' points on the network are less renormalized and lie closer to the physical degrees of freedom located on the boundary (input layer). Moreover the bulk degrees of freedom are regarded as holographic in the sense that their dynamics can be described by a theory purely defined using the boundary degrees of freedom. It is through this holographic correspondence that a tantalizing analogy to quantum gravity is often made. This analogy which is motivated by ideas from AdS/CFT holography has been shown to be robust.\cite{swingle2012entanglement,molina2015information,van2009comments,van2010building,swingle2010mutual,molina2011holographic,molina2011connecting,balasubramanian2012momentum,matsueda2011scaling,ishihara2013refined} For example explicit identifications with Functional Renormalization Group (FRG) and emergent gravity have recently been made.\cite{Leigh_Onkar_Alex_spinholo,douglas2011holographic}. 

Despite the deep connection to quantum gravity/holography, MERA and its myriad generalizations\cite{qi2013exact,mollabashi2014holographic,haegeman2013entanglement,miyaji2015cmera,molina2015information,konig2009exact,gu2016holographic,wen2016holographic} are still regarded as highly evolved numerical algorithms to simulate ground states. Past computational studies have mostly concentrated on CFT or near CFT ground states and Gibbs thermal ensembles with an emphasis on highly accurate and scalable quantum simulation. \cite{vidal2007entanglement,Vidal&friends_scaling_Entrenom,vidal&evenvbly_tensornet_MERA} Less emphasis has been placed in the RG interpretation. Even though the RG viewpoint has provided a conceptual framework to make the holographic correspondence concrete. 

In this work, we employ the MERA tensors obtained by the original RG procedure as described by Vidal,\cite{vidal2007entanglement,vidal&evenbly_MERA_QC} as an \emph{analysis tool} to study the low energy dynamics of the 1D transverse {field Ising model} in the paramagnetic phase. This is a very natural enterprise given that the MERA tensors literally define a quantum circuit.\footnote{More generally the family of tensor networks which includes MERA, in principle defines a quantum circuit diagram which is not necessarily planar. The category of all such quantum circuits are either in the concrete category {\bf FinHib} of finite dimensional Hilbert spaces or specializations thereof [See references \onlinecite{coecke2010categories,selinger2010survey}].} And as a quantum circuit, it can act on any pure state wavefunction, much like a quantum Fourier transform. As we shall demonstrate with the 1D transverse field Ising model, if the physical ground state has an accurate representation as a MERA circuit acting on a trivial product state in the bulk, then we will have obtained a reference vacuum for which to \emph{compare} low energy excited states against. 
This leads us to a new definition of a stable holographic quasiparticle which we call the ``\emph{hologron}" or holographic quantum excitation. 
By performing a unitary transformation on the physical Hamiltonian using the MERA quantum circuit, 
we determine the bulk Hamiltonian in terms of these holographic degrees of freedom. 
A main result of this work, is that the 1-hologron sector or single quasiparticle state is stable in the gapped paramagnetic phase of the transverse field Ising model and accurately describes the lowest excited state subspace. Moreover, we demonstrate with the MERA quantum circuit that hologrons are dual to extended physical excitations on the boundary. Thus we demonstrate that MERA can efficiently realize the holographic correspondence away from the CFT point and the duality between bulk and boundary in this instance is a ``\emph{correlated-uncorrelated}" duality between the physical ground and the bulk ground state. Specifically, the bulk ground state is uncorrelated or weakly entangled, but at the cost of a more complicated bulk Hamiltonian residing in one dimension higher. 

Lastly, we should mention that a connection between MERA and wavelet transforms has recently been established.\cite{lee2016exact,evenbly2016entanglement,brennen2015multiscale,singh2016holographic} Whereby MERA or the simpler Exact Holographic Mapping (EHM)\cite{qi2013exact} can be regarded as a quantum circuit realization of an orthonormal wavelet transform acting on the mode space of the physical fermionic degrees of freedom. These fermions are the Jordan-Wigner transformed physical spin-1/2 degrees of freedom. These works, that are focused on  the CFT point of the transverse field Ising model, can yield simple analytic forms for the tensors and simple relationships between the bulk and boundary fermions. In this paper however, we will be almost entirely focused on the massive phase of the transverse field Ising model where correlations are gapped and the ground state in the bulk strongly approximates an exact tensor product state, or trivial vacuum.  

This paper is organized as follows. In Section \ref{sec:formulation} we describe the specific MERA tensor network that we have constructed and discuss the unitary quantum circuit that it defines. Then in Section \ref{sec:MERA_and_hologrons}, we delve into the uses of a well optimized MERA network which is then used to define the bulk hologrons. Next in Section \ref{sec:TI_model}, we describe the transverse field Ising model and briefly touch upon its phases and present real-time dynamical data of a locally excited paramagnetic ground state. Following that in Section \ref{sec:bulk_MERA_dynamics}, we then use the MERA quantum circuit to holographically analyze the real-time dynamics. This then leads to Section \ref{sec:HBulk} where we study in detail the effective bulk dynamics that is obtained by the MERA quantum circuit. This section is the main part of the paper and discusses in detail all the ingredients that go into making the holographic correspondence between hologrons and physical spin-flips. Then in Section \ref{sec:emergent_geometry} we present our speculations about how a bulk geometry might emerge from taking the continuum limit of the 1D quantum spin chain. Finally we conclude the paper in Section \ref{sec:conclusions}, where we summarize our main findings and present future speculations regarding MERA. 

\section{The MERA Quantum Circuit}\label{sec:formulation}

In this section, we introduce our notation for the MERA. We shall limit our study to spin-1/2 chains as our model system and only consider periodic chains. Let $\mathcal{H}_{s}$ be the physical Hilbert space of $L$ spins denoted by ${s}_{i}$, and let $\mathcal{H}_t$ be an auxiliary or \emph{ancillary} Hilbert space of $L$ spins which we refer to as the bulk qubits and are denoted by ${t}_\mu$. Both these spaces are of dimension $2^L$ and in the holographic correspondence the former is identified with the boundary (physical) degrees of freedom while the latter with the bulk (holographic) degrees of freedom. 
\begin{figure*}
	\includegraphics[width=\textwidth]{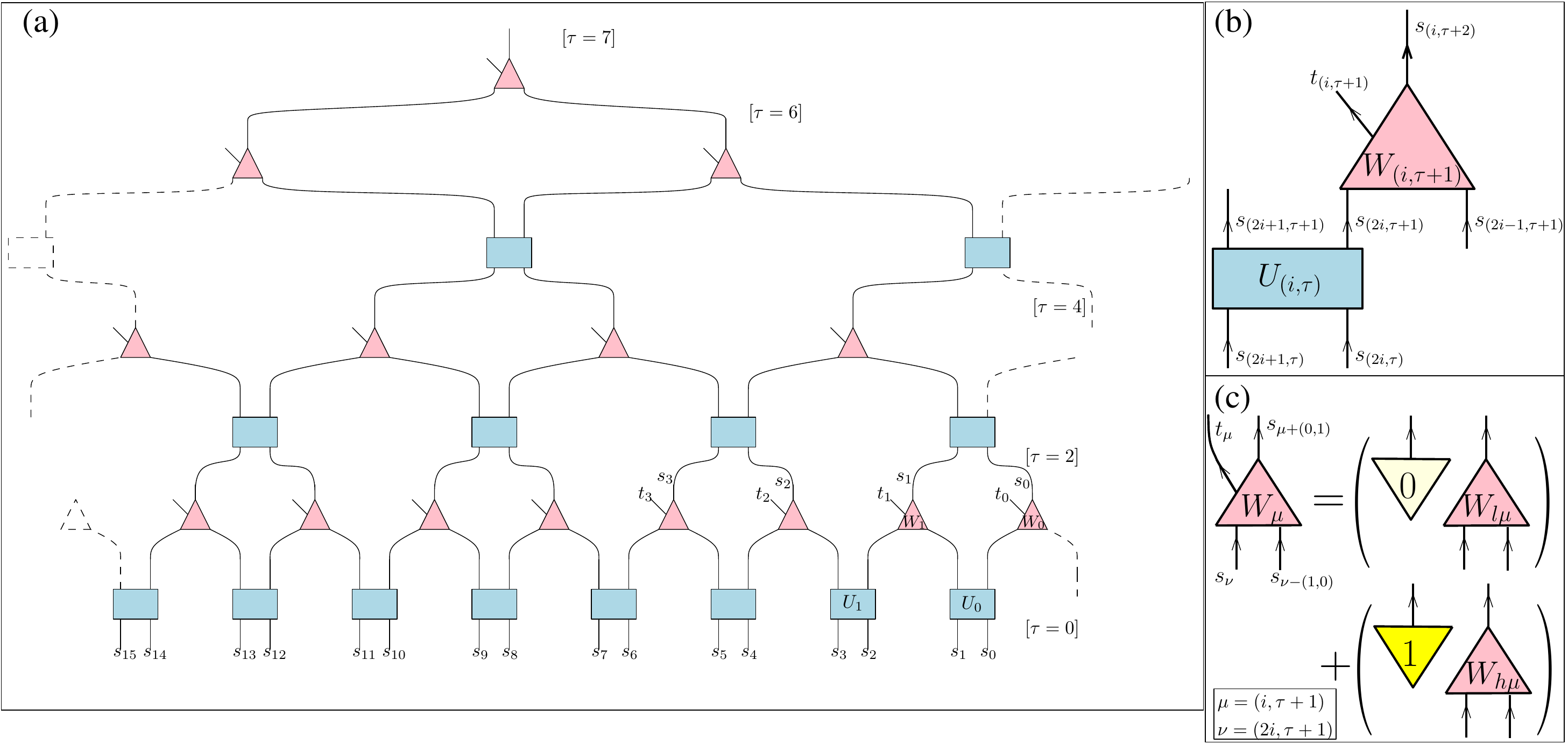}
	\caption{(color online) (a) A MERA quantum circuit defined on a graph $\Gamma$ with periodic boundary conditions and 16 inputs $\{ s_{(0,\tau=0)},\ldots, s_{(15,\tau=0)} \}$ at the bottom. The flow of ``time" $[\tau]$ of the circuit is from bottom to top. At the nodes of the graph are disentanglers (blue rectangles $\color{lightblue} \blacksquare$) and isometries (red triangles $\color{pink} \blacktriangle$). Located at the isometries are hanging edges which we identify as the bulk qubits $\{ t_{(i,\tau)} \}_{(i,\tau)}$. They are the outputs of the MERA circuit computation. The dashed lines and tensors denote connections that go around the periodic graph. On the lower right corner are some labeled tensors and edges (qubits) where we have suppressed the time index for clarity. Note that at the highest level or latest $[\tau]$, there is no need for a final disentangler gate. (b) The basic unit of a MERA network from which the entire network is constructed, where the edges are decorated with arrows to emphasize the direction of information flow. Our convention for the site ordering between adjacent $\tau$ layers is also described in this basic unit. (c) The decomposition of the ``isometry"  $W_\mu$ into low $W_{l\mu}$ and high $W_{h\mu}$ isometries.}
\label{fig:MERA_network}
\end{figure*}
Another perspective which is the computational viewpoint -- which we will take -- is that the boundary degrees of freedom are logical inputs to the logical outputs that are the bulk degrees of freedom for the MERA quantum circuit.

The MERA which we denote by $\bf {M}$ is a finite depth \emph{unitary} quantum circuit acting between the physical (boundary) and holographic (bulk) Hilbert spaces 
\begin{align}
	{\bf M} : \mathcal{H}_\text{s} \rightarrow \mathcal{H}_\text{t}
\end{align}
defined by two collections of SU(4) special unitary matrices. They are customarily known as \emph{isometries} $\{W_{(i,\tau)}\}$ and \emph{disentanglers} $\{ U_{(i,\tau)}\} $. The labels ${(i,\tau)}$ these tensors carry are coordinates of nodes on a graph $\Gamma$ which resembles a tree and is shown in Fig.{\ref{fig:MERA_network}(a,b)}. Here $\tau$ is the level, layer or height and $i$ is a position along the horizontal direction. The nodes of $\Gamma$ will be denoted by Greek letters like $\mu=(i,\tau)$ to emphasize their discrete space-time interpretation. Also we will continue to refer to the $W_\mu$ tensors as isometries even though they are SU(4) valued like the disentanglers $U_\mu$. On the edges of $\Gamma$ live ``world-lines" of local qubits or spin-1/2 degrees of freedom ${s}_{(i,\tau)}=0,1$. They are also labeled by space-time indices in the convention of Fig.\ref{fig:MERA_network}(b), and are acted on by the isometries and disentanglers at the nodes of $\Gamma$. At the lowest level $\tau=0$ the ${s}_{(i,0)}$ spins are identified with the degrees of freedom in $\mathcal{H}_s$ and serve as inputs for $\MM$. 

In this light, the space $\mathcal{H}_s$ is to be regarded as a computational space that is acted on by the gates of $\bf M$. While $\mathcal{H}_t$ is a coding space which is the output of quantum circuit defined by $\MM$. This viewpoint has recently provided a new conceptual framework to understand MERA and AdS/CFT holography in terms of quantum error correction codes.\cite{Preskill_and_gang_holo_quant_error_code} It is easily appreciated that under $\MM^\dagger$ the bulk spins ${t}_\mu$ are highly non-local degrees of freedom in $\mathcal{H}_s$. 

In general the MERA network is a highly customizable quantum circuit and there exists now a great many variations and generalizations.\cite{vidal_MERA_2D,Lukasz_QIsing_MERA_2D,vidal2009entanglement,vidal_MERA_2D,evenbly2009algorithms,corboz2010simulation,evenbly2014class} Depending on the application, the graph $\Gamma$ which defines the topology and connectivity of the circuit may be selected differently. The $U_\mu$ and $W_\mu$ tensors may also come in different shapes and ranks, so long as they are compatible with $\Gamma$. In fact, in the original MERA\cite{vidal2007entanglement} the isometries $W_\mu$ are named so because they do not possess the full rank (surjectivity) but are 1 to 1 (injectivity) in the sense that $W_\mu W^\dagger_\mu=\mathbbm{1}$ but $W^\dagger_\mu W_\mu \neq \mathbbm{1}$. Another common variation is the dimension of an internal edge, which is known as the bond dimension $\chi$ is often not fixed to that of a single qubit ($\chi=2$) in more conventional MERA realizations. Often $\chi$ is made to be as large as is necessary to obtain an optimally acceptable variational ground state energy. 

\section{MERA and the hologrons}\label{sec:MERA_and_hologrons}

Having introduced MERA and the specific realization that we shall use, we next discuss its uses as a `holographic transform' to define bulk degrees of freedoms which we call hologrons. The task of determining the disentangler and isometry tensors from any given ground state wavefunction $|\Omega_0\rangle$ is computationally involved and we have relegated the relevant details of our methods to Appendix \ref{app:MethodDisIso}. We remark however that our method follows closely Vidal's original scheme\cite{vidal2007entanglement} where local disentanglers are chosen to minimize the entanglement entropy between a contiguous block of 2 sites and its environment. 

The conceptual novelty of MERA due to Vidal,\cite{vidal2004} lies in its choice of tensor network that mimics a dynamical process in a discrete space-time. In the MERA tensor network, local quantum gates act on a scale by scale basis to remove short range entanglement (by disentanglers) and to coarse-grain degrees of freedom (by isometries) when interpreted in the increasing RG $\tau$-direction, viz. ${\bf M} : \mathcal{H}_s \rightarrow \mathcal{H}_t$. Dually, when read in the ${\bf M}^\dagger : \mathcal{H}_t \rightarrow \mathcal{H}_s$ direction,\cite{swingle2010mutual} the RG process in reverse acts to introduce short range entanglement (by disentanglers) and expand the degrees of freedom (by isometries) on a scale by scale basis, such that final outcome will contain sufficient complexity to be a viable ansatz wavefunction. It is expected that these mutually dual descriptions perform best when the system at hand is a critical CFT, where scale invariance is reflected by self-similar tensors of the MERA network (infinite sized graphs) and is seen to be an extremely efficient representation of critical ground states.\cite{vidal&evenbly_MERA_QC,vidal_MERA_2D,evenbly2016entanglement} 

For our purposes we shall appeal more to the RG interpretation of MERA, where we seek to optimize the tensors on the fixed graph $\Gamma$ defined in Fig.\ref{fig:MERA_network} such that for a given ground state $|\Omega_0\rangle$, we maximize the overlap  between the ansatz and the true ground state 
\begin{align}
|\langle 0 | \MM | \Omega_0 \rangle |^2
= |\langle \Omega(\MM)|\Omega_0\rangle|^2 \leqslant 1
,\label{eqn:fidel}
\end{align}
where $|0\rangle = \underset{\mu}{\bigotimes} |0_{\mu}\rangle \in \mathcal{H}_t$ is the completely 0-polarized product state of the bulk qubits ${t}_{\mu}$ each labeled by a space-time index $\mu=(i,\tau)$. Here $|\Omega(\MM)\rangle := \MM^\dagger |0_t\rangle \in \mathcal{H}_s$ is the MERA ansatz which is the state prepared from $|0\rangle$ by $\MM^\dagger$. The overlap condition  amounts to maximizing the fidelity of $|\Omega(\MM)\rangle$ with $|\Omega_0\rangle$. If $|\Omega_0\rangle$ is the exact ground state of the Hamiltonian $H$ which saturates the minimum energy bound, then an equivalent criterion is the minimization of variational energy $E_\text{var}(\MM)=\langle \Omega(\MM) | H |\Omega(\MM)\rangle$.

Contrary to the commonly used form of MERA where $\MM$ is norm non-increasing, we have taken $\MM$ to be unitary without coarse-graining, such that it describes a \emph{reversible} quantum circuit. Our motivation for doing so is to allow the bulk qubits ${t}_\mu$ to become more than just bulk labels for the physical $|\Omega_0\rangle$ by allowing them to acquire quantum fluctuations. A related unitary MERA such as this was previously proposed and is known as the Exact Holographic Mapping (EHM). \cite{qi2013exact} In the EHM network the quantum fluctuating bulk qubits ${t}_\mu$ characterize the emergent quantum geometry of the AdS dual in the holographic correspondence. The direction of this work is very much inspired by the EHM proposal. However our circuit differs with EHM in regard to the disentanglers which are entirely neglected in EHM. This is because disentanglers are absolutely necessary for an accurate representation of the ground state as a trivial bulk product state in the $t_\mu$ basis.

Next by allowing the ${t}_\mu$ bulk degrees of freedom to fluctuate, we can attach physical significance to the states with ${t}_\mu = 1$, but only when optimal disentanglers  -- with respect to $|\Omega_0\rangle$ -- are included in $\MM$. In particular, such states describe physical wavefunctions that are strongly orthogonal to $|\Omega_0\rangle$ in the sense that
\begin{align}
\langle \Omega_0|\MM^\dagger \prod_{\mu \in \mathcal{M}} \chi^x_\mu |0\rangle \approx 0 \label{eqn:t_ortho}
\end{align}
when $\MM$ is optimal in sense of maximizing the LHS of (\ref{eqn:fidel}). Here the operators $\chi^x_\mu \equiv {X}_\mu$, $\chi^y_\mu \equiv {Y}_\mu$, $\chi^z_\mu \equiv {Z}_\mu$ denote the local Pauli gates acting on the ${t}_\mu$ qubit and $\mathcal{M}\subset \Gamma$ is any non-empty collection of bulk qubits. The strong orthogonality statement is clear since the states $\chi^x_\mu|0_\mu\rangle$ and $|0_\mu\rangle$ are orthogonal. More precisely for any $|\psi_t\rangle\in \mathcal{H}_t$ orthogonal to $|0\rangle$ one has from the Cauchy-Schwartz inequality
\begin{align}
|\langle  \Omega_0|\psi_{\MM}\rangle| 
&= \;|\langle \Omega_0-\Omega(\MM)|\psi_{\MM} \rangle| \nonumber \\
&\leqslant \;\parallel \Omega_0-\Omega(\MM) \parallel \cdot \parallel \psi_{\MM} \parallel \nonumber \\
&= \;2^{1/2}\sqrt{1-\text{Re}\langle \Omega(\MM)|\Omega_0\rangle} \parallel \psi_{\MM}\parallel
\end{align}
where $|\psi_{\MM}\rangle = \MM^\dagger |\psi_t\rangle \in \mathcal{H}_s$. Hence (\ref{eqn:fidel}) acts as to maximally bound the overlap. Moreover, assuming that $|\Omega_0\rangle$ is a ground state with minimal energy and has a strong overlap with $\MM^\dagger|0_t\rangle$, then any $\MM^\dagger\prod_{\mu \in \mathcal{M}} {\chi^x_\mu} |0\rangle$ must be an excited state with greater average energy. However, these excited states cannot be interpreted as energy eigenstates since there is no guarantee that $\MM^\dagger\prod_{\mu \in \mathcal{M}}\chi^x_\mu |0\rangle$ produces an eigenstate of $H$. One is only sure that non-local energetic perturbations of $|\Omega_0\rangle$ are produced by the action of $\chi^x_\mu$ in the bulk. From the tree structure of Fig.\ref{fig:MERA_network}(a), the degree of non-locality is strictly ordered by the $\tau$ coordinate in $\Gamma$ with the more physically non-local ${t}_\mu$'s possessing higher values of $\tau$. 

The set of states $\prod_{\mu \in \mathcal{M}}\chi^x_\mu |0\rangle$ with $\mathcal{M}$ arbitrary form a complete orthonormal basis set in $\mathcal{H}_t$ and it is convenient to label them by occupation numbers 
\[
\nn = \{n_\mu \in \{0,1\} \text{ for all }\mu \} 
\]
with respect to ${t}_\mu \equiv (\chi^z_\mu + \mathbbm{1})/2$. Concisely, we define the orthonormal basis states
\begin{align}
|\nn\rangle := \prod_{\mu} (\chi^x_\mu)^{n_\mu}|0\rangle.
\end{align}
Then we take the states with $t_\mu=1$ as a definition of a \emph{hologron} quasiparticle in the bulk with $|0\rangle$ as their number vacuum and $\{|\nn \rangle\}$ as the set of hologron occupation configuration states. Being qubits, the newly defined hologrons are neither bosonic nor are they fermionic particles.\footnote{More generally, they could be re-interpreted as hard core bosons or alternatively as Jordan-Wigner fermions in the bulk.} Furthermore, we can define a \emph{total hologron number} by
\begin{align}
N_t := \sum_{\mu} t_\mu  
\end{align}
where the unique MERA ansatz ground state $|0\rangle$ occupies the zero number sector. It should be emphasized that the hologrons are localized qubits in the bulk and that a $\mathbb{Z}$ grading of $\mathcal{H}_t$ by the eigenspaces of $N_t$ is not only convenient, but as we shall see in Sec.\ref{sec:HBulk} is useful for developing effective low energy holographic Hamiltonians that act in the bulk. 

Roughly speaking, a wavefunction with more hologrons is more highly excited, and projecting $\mathcal{H}_t$ onto a small number of hologrons can provide a low energy effective subspace.  Nevertheless, a more physical interpretation of hologron excitations can be gained from analyzing the procedure used to construct the isometries $W_\mu$ which is described in more detail in Appendix \ref{app:MethodDisIso}. We remark that in Ref.\onlinecite{oberreuter2015representation} where an analytical exact MERA is known for the dual network of the 2D toric code model, similar interpretations are reached regarding the quantum information of excited states for that model.  

\begin{figure}
	\includegraphics[width=0.35\textwidth]{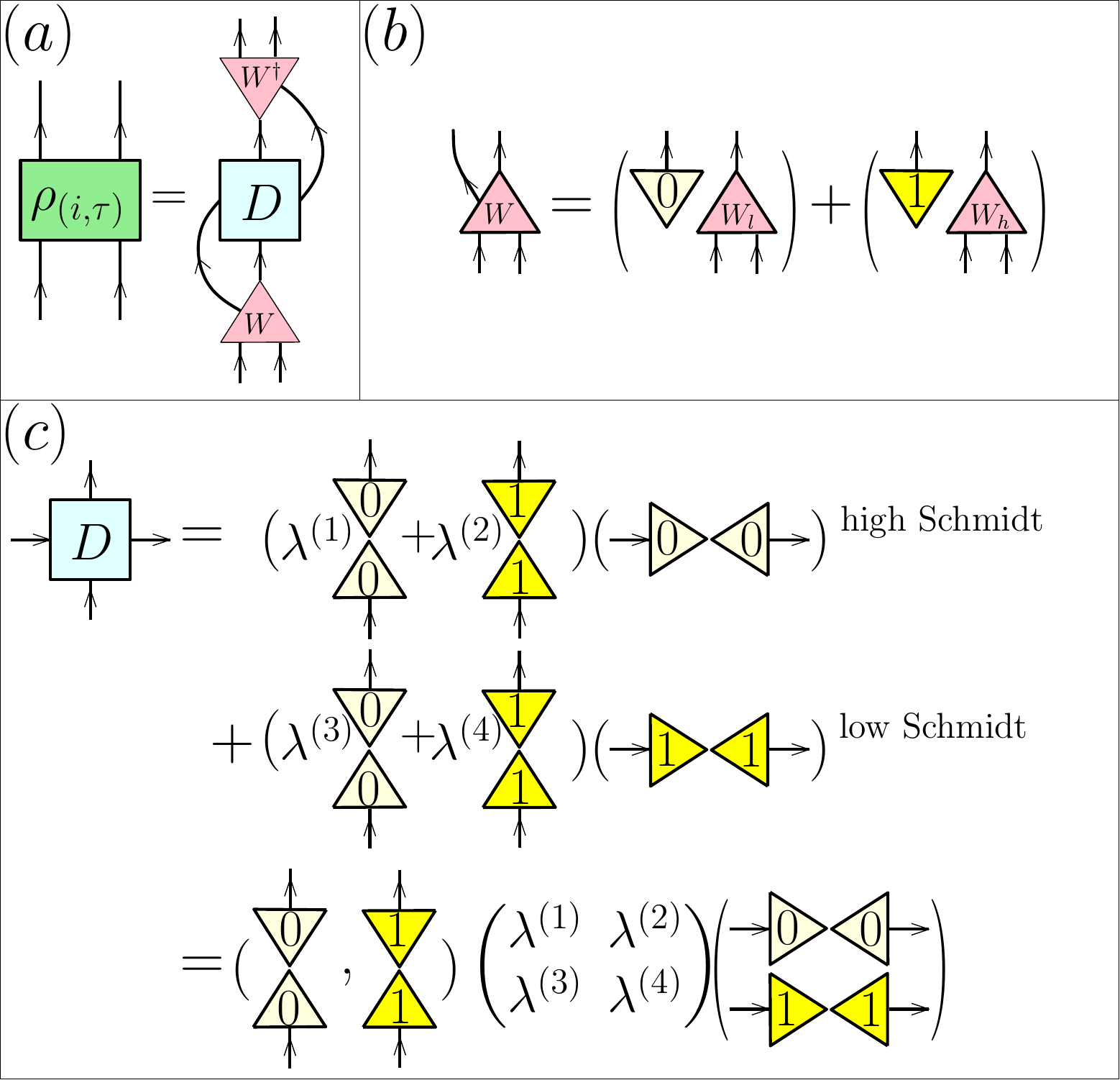}
	\caption{(color online) (a) The decomposition of the disentangled 2-site reduced density matrix $\rho_{(i,\tau)}$ by the $W$ unitary tensor. (b) The $W$ tensor is further decomposed into the low and high isometries $W_{l,h}$. (c) The tensor $D$ is a diagonal matrix with entries of the eigenvalues of $\rho_u^A$ ordered by the $|s,t\rangle =|0,0\rangle,|0,1\rangle,|1,0\rangle,|1,1\rangle$ states. The $t=0$ states correspond to the low $W_l$ isometry with the two largest Schmidt eigenvalues $\lambda^{(1,2)}$ and the converse for $t=1$. }
	\label{fig:app_rho}
\end{figure}

First, recall the iterative procedure to construct a MERA network from entanglement renormalization. Consider the application of MERA quantum circuit with coarse graining on $|\Omega_0\rangle$ up to an even $\tau$-layer. This, in the entanglement RG picture produces a new pure state wavefunction that has been renormalized. Then to determine the next layer's tensors we first trace out a contiguous block of 4-spins where disentanglers are to be applied. Now after applying the pair of disentanglers gates (assuming that they have been optimized) to this 4-site block, a partial trace is taken over the sites on the boundaries which produces a 2-site reduced density matrix $\rho_{(i,\tau)}$. The eigenvalue decomposition of this block reduced density matrix $[\rho_{(i,\tau)}]_{s_{(2i,\tau)}\, s_{(2i-1,\tau)}}^{\bar{s}_{(2i,\tau)}\, \bar{s}_{(2i-1,\tau)}}$ yields four positive semi-definite eigenvalues $\lambda^{(1)}_{(i,\tau)} \geq \ldots \geq \lambda^{(4)}_{(i,\tau)} $ and the their associated (Schmidt) eigenvectors $|\lambda^{(1)}_{(i,\tau)}\rangle ,\ldots,|\lambda^{(4)}_{(i,\tau) } \rangle$. The set of 4 eigenpairs can be divided into two sets of large $\{\lambda^{(1)}_{(i,\tau)}, \lambda^{(2)}_{(i,\tau)}\}$ and small $\{\lambda^{(3)}_{(i,\tau)}, \lambda^{(4)}_{(i,\tau)} \}$ eigenvalues.  Correspondingly in an entanglement spectrum they are the low and high entanglement energy states respectively. In the usual MERA where the isometries are coarse-graining in accordance to White's rule,\cite{vidal2009entanglement} $W_{(i,\tau)}$ is defined as the tensor that \emph{projects} onto the high $\lambda^{(n)}_{(i,\tau)}$ (low entanglement energy) subspace. This {defines} a new $s$-qubit in the next $\tau-$time from these eigenstates. We call this tensor the low isometry $W_{l(i,\tau)}$ and define it precisely as 
\begin{align}
W_{l(i,\tau)}:= |{s}_{(i,\tau+1)}=0 \rangle \langle \lambda^{(1)}_{(i,\tau)}| 
+|{s}_{(i,\tau+1)}=1\rangle \langle \lambda^{(2)}_{(i,\tau)}|. 
\end{align}
Note that the output states are to be identified with the $s$-states for the next $(\tau+1)$ MERA time layer. As such, the low isometry has the index structure $[W_{l(i,\tau)}]_{s_{(i,\tau+1)}}^{s_{(2i,\tau)}\,s_{(2i-1,\tau)}}$. One should think of the new $s-$qubit as a handle for one of the low energy states $\lambda^{(1)}_{(i,\tau)}$ or $\lambda^{(2)}_{(i,\tau)}$, that carries the two most dominant correlations or largest support of $\rho_{(i,\tau)}$. By contrast, for the least two dominant contributions, we define the high isometry $W_{h(i,\tau)}$ in an entirely complementary manner as
\begin{align}
W_{h(i,\tau)}:= |{s}_{(i,\tau+1)}=0\rangle \langle \lambda^{(3)}_{(i,\tau)}| 
+|{s}_{(i,\tau+1)}=1\rangle \langle \lambda^{(4)}_{(i,\tau)}|
\end{align}
which has the same index structure as $W_{l(i,\tau)}$ and so has the same types of inputs and outputs. Unless the support of $\rho_{(i,\tau)}$ is exactly in low energy subspace, the high isometry tensor $W_{h(i,\tau)}$ plays a role in encoding some of the correlations of $\rho_{(i,\tau)}$; as residual as they may be. The hologron qubit ${t}_{(i,\tau+1)}$ is then defined as the degree of freedom that determines which entanglement energy subspace is propagated into the next $(\tau+1)$ layer. That is whether on a \emph{low} setting $t_{(i,\tau)}=0$ which implies $W_{l(i,\tau)}$ or a \emph{high} setting $t_{(i,\tau)}=1$ which implies $W_{h(i,\tau)}$. This can be made precise by defining the unitary tensor $W_{(i,\tau)}$ as
\begin{align}
W_{(i,\tau)} :=&\hspace{0.4cm}
|{t}_{(i,\tau+1)}=0\rangle \otimes W_{l(i,\tau)} \nonumber\\
&+|{t}_{(i,\tau+1)}=1\rangle \otimes W_{h(i,\tau)} \label{eqn:W_decomp}.
\end{align}
In terms of tensor network diagrams this decomposition of $\rho_{(i,\tau)}$ and $W_{(i,\tau)}$ in (\ref{eqn:W_decomp}) is shown in Fig.\ref{fig:app_rho}.  Hence, intuitively the presence of a hologron $t_\mu=1$ localized at  $W_\mu$ in the bulk represents the occupancy of the high entanglement energy subspace. Conventional MERA with coarse-graining selects only the low isometries and is equivalent to a post-selection measurement into $t_\mu=0$ for all $\mu$ in the bulk. 

Finally, we should remark that the hologron states are very much analogous to the bulk excitations that have been recently identified within cMERA,\cite{miyaji2015cmera} the exact FRG\cite{fliss2016unitary} and AdS/CFT formalisms\cite{verlinde2015poking}. However a key difference being that the hologrons are in exact correspondence with the boundary Hilbert spaces, given that they are defined using a unitary transformation much like the Exact Holographic Method.\cite{qi2013exact} Also we should re-emphasize that they are distinguishable quantum particles described by qubit degrees of freedom.

\section{The Transverse Field Ising model}\label{sec:TI_model}

\subsection{Hamiltonian and Phases}

It must be borne in mind however that whether or not the fidelity optimality criterion (\ref{eqn:fidel}) is attainable depends strongly on the model Hamiltonian at hand and the chosen MERA graph $\Gamma$. Nevertheless, there has been some precedent in quantum spin-chains and in particular the transverse field Ising model where efficient distentangling has been achieved\cite{vidal2007entanglement}, especially in its gapped phases. The 1D transverse field Ising model Hamiltonian is defined as 
\begin{align}
H = - \sum_{i=0}^{L-1} {\sigma}^x_{i+1} {\sigma}^x_{i} + h^z \sum_{j=0}^{L-1} {\sigma}^z_j
\end{align}  
where the transverse field $h^z$ is a real coupling and $({\sigma}^z_i+\mathbbm{1})/2$ is identified with ${s}_{(i,0)}$ which are the input qubits for $\MM$. The transverse field Ising model is exactly solvable by the Jordan-Wigner transformation and exhibits two phases as a function of $h^z$; both of which are gapped. The critical point at $|h^z|=1$ is gapless and is described by a $c=1/2$ CFT in the continuum limit. For the $|h^z|>1$ paramagnetic phase, the ground state is non-degenerate, while the $|h^z|<1$ ferromagnetic phase is doubly degenerate and spontaneously breaks the global Ising symmetry ${\sigma}_i^x \rightarrow - {\sigma}_i^x$ in the thermodynamic limit. 
Strictly speaking, either gapped phase is conventionally ordered with ordering in $\sigma^z_i$ in the paramagnetic phase and ordering in $\sigma^x_i$ in the ferromagnetic phase. But it is only the latter that is spontaneous symmetry breaking. 

In the rest of the paper, we will limit our discussion to the paramagnetic phase with $h^z=3$ in an $L=16$ chain with periodic boundary conditions, primarily because this phase has a non-degenerate classically ordered ground state that is well approximated by our MERA ansatz.

\subsection{Real-time dynamics}\label{sec:real_dynamics}

\begin{figure*}
	\frame{\includegraphics[scale=0.45,valign=t]{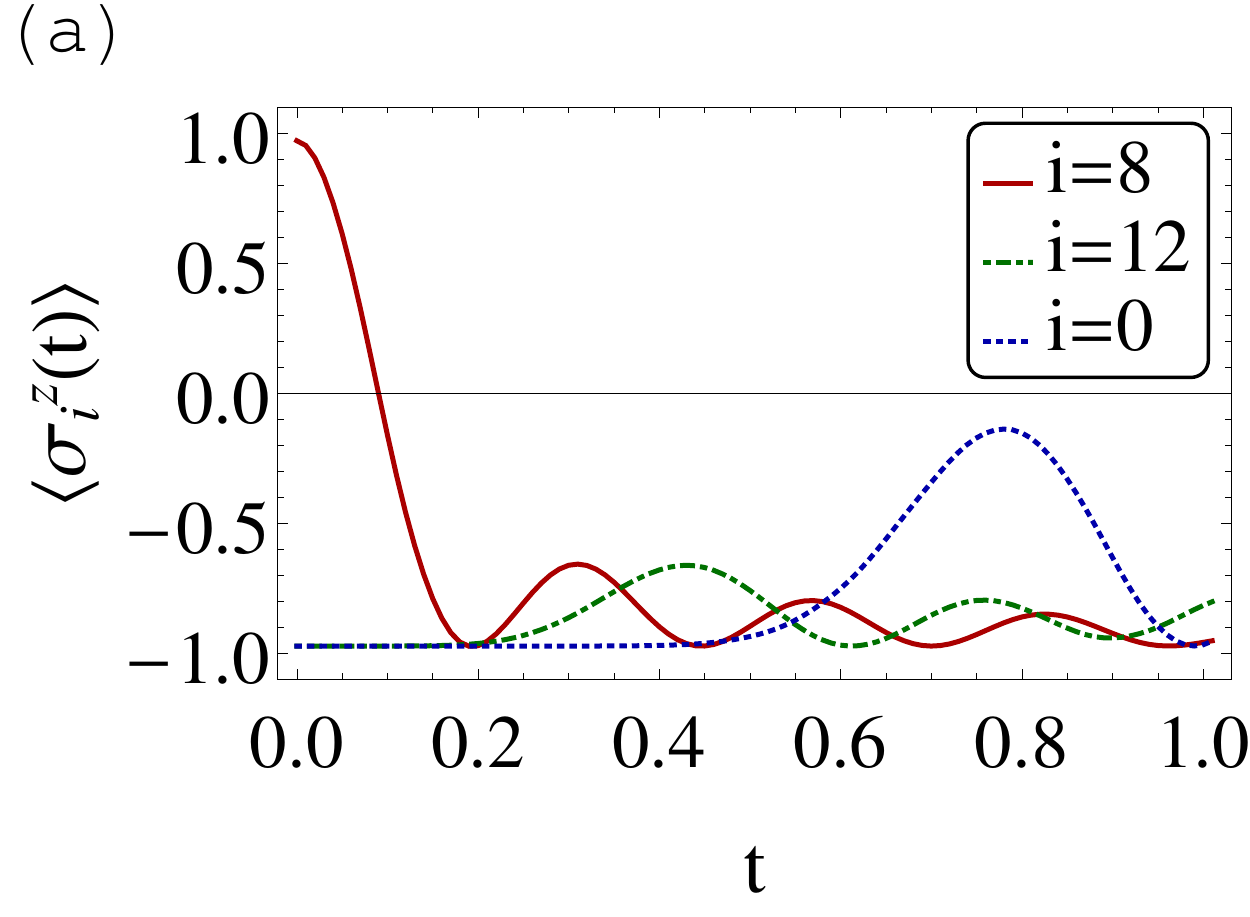}
	\includegraphics[scale=0.45,valign=t]{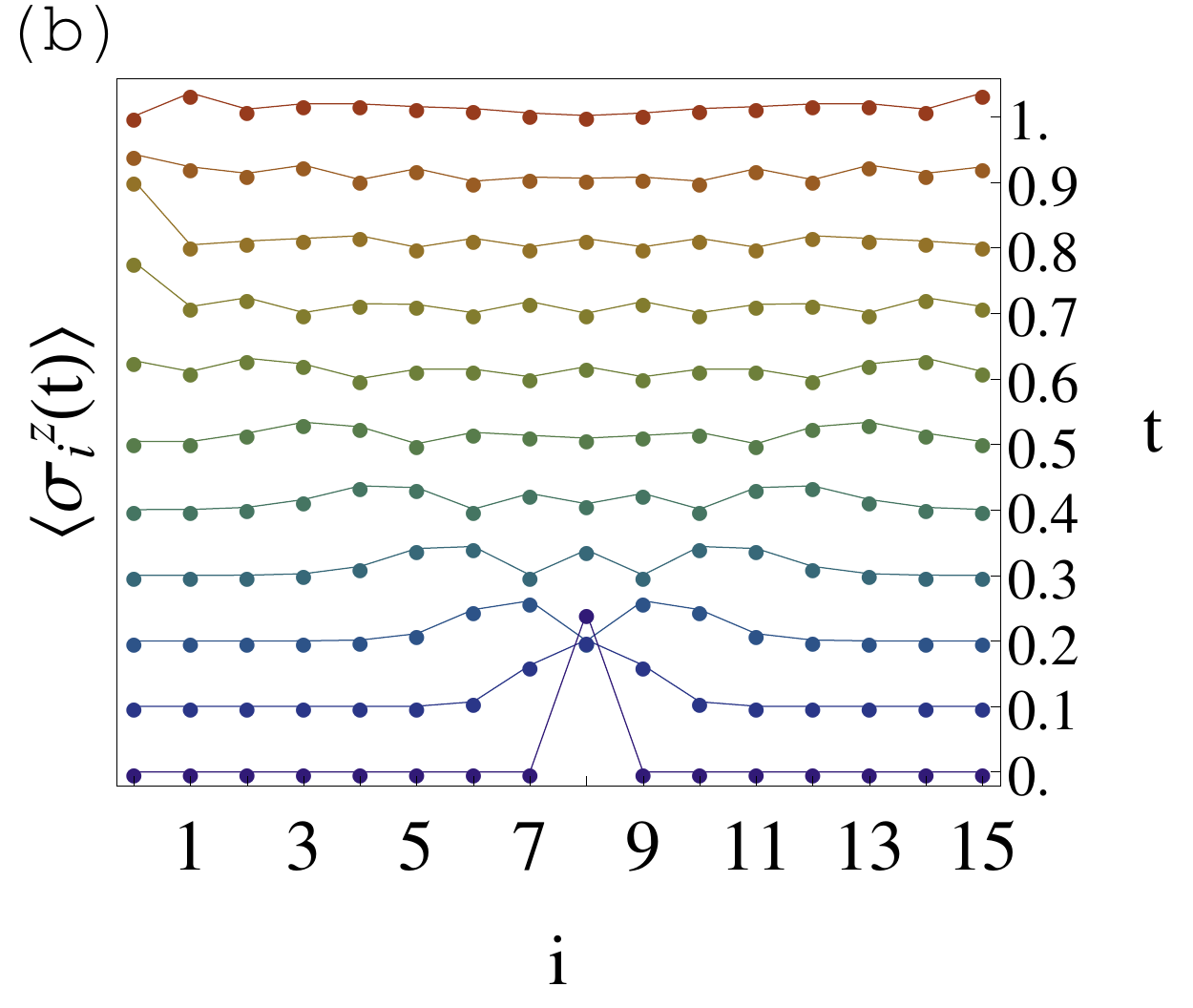}\;
	\includegraphics[scale=0.4,valign=t]{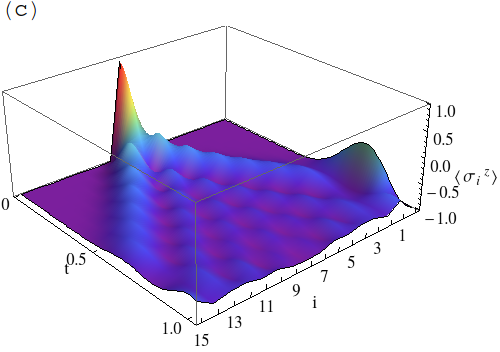}}
	\caption{(color online) The time development of a locally excited ground state in the paramagnetic phase with $h^z=3$ in a periodic chain of length $L=16$. The initial ground state uniform magnetization is $\langle \sigma_i^z\rangle\approx -0.972$. The system is excited by flipping a spin at position $i=8$ at time $\st =0$. (a) Time development  of $\langle \sigma_i^z(\st)\rangle$ at selected sites $i=0,8,12$. (b) Variation of $\langle \sigma_i^z(\st)\rangle$ in position at shifted times. (c) Overall time evolution of $\langle \sigma_i^z(\st)\rangle$ as a 2D surface by interpolating between discrete sites $i$ of the chain.}
\label{fig:Sztime}\end{figure*}

Next we consider the real-time evolution of a low energy excited state of the transverse field Ising model. 
From the discussion in Sec.\ref{sec:MERA_and_hologrons}, one can expect that the low energy excited states will appear non-trivially under the MERA unitary transform. Furthermore, by studying their real-time dynamics with the MERA transform, we can analyze the spread of information over time in a novel way with the bulk degrees of freedom that are naturally organized according to correlation length scale. We set out to achieve this in a numerical experiment by exciting the ground state of $H$ by locally flipping a spin. The excited state wavefunction is then time-evolved using the unitary operator  $\e^{-i 2\pi H \text{t}}$ and it is observed that the excess energy de-localizes over time. A main goal of this paper is then to understand how this energy diffusion progresses from the holographic point of view by using the MERA quantum circuit $\MM$.

Technically, we obtain the $L=16$ numerical ground state by exact diagonalization using the Lanczos method.\cite{lehoucq1998arpack} Then injecting energy into the system involves a simple local  $\sigma^x_i$ operation on the numerical ground state $|\Omega_0\rangle$. Time evolution of the many-spin wavefunction is then carried out using a first order implicit Crank-Nicholson method\cite{press2007numerical} with a time step of $\delta \st = 10^{-3}$. The necessary linear sparse matrix inversions operations were carried out with the UMFPACK\cite{UMFPACKDavis2004} library.

Shown in Fig.\ref{fig:Sztime} are results of a time-evolution simulation for $L=16$ and $h^z= 3$ after a spin-flip at position $i=8$. From the time development of the $\langle \sigma^z_i(\text{t})\rangle$ profile,\footnote{The time-dependent operator here is understood to mean the Heisenberg picture time-dependent operator under the full Hamiltonian $H$.} it is clear that the excited state is out of equilibrium and is evolving ballistically. Waves of $\langle \sigma^z_i(\text{t})\rangle$ are observed and their wavefronts propagate at finite speeds within the Lieb-Robinson bound\cite{lieb2004finite} which can be determined from the exact solution of the model. 

\section{Bulk Dynamics from MERA}\label{sec:bulk_MERA_dynamics}

\begin{figure*}
	\includegraphics[width=0.9\textwidth]{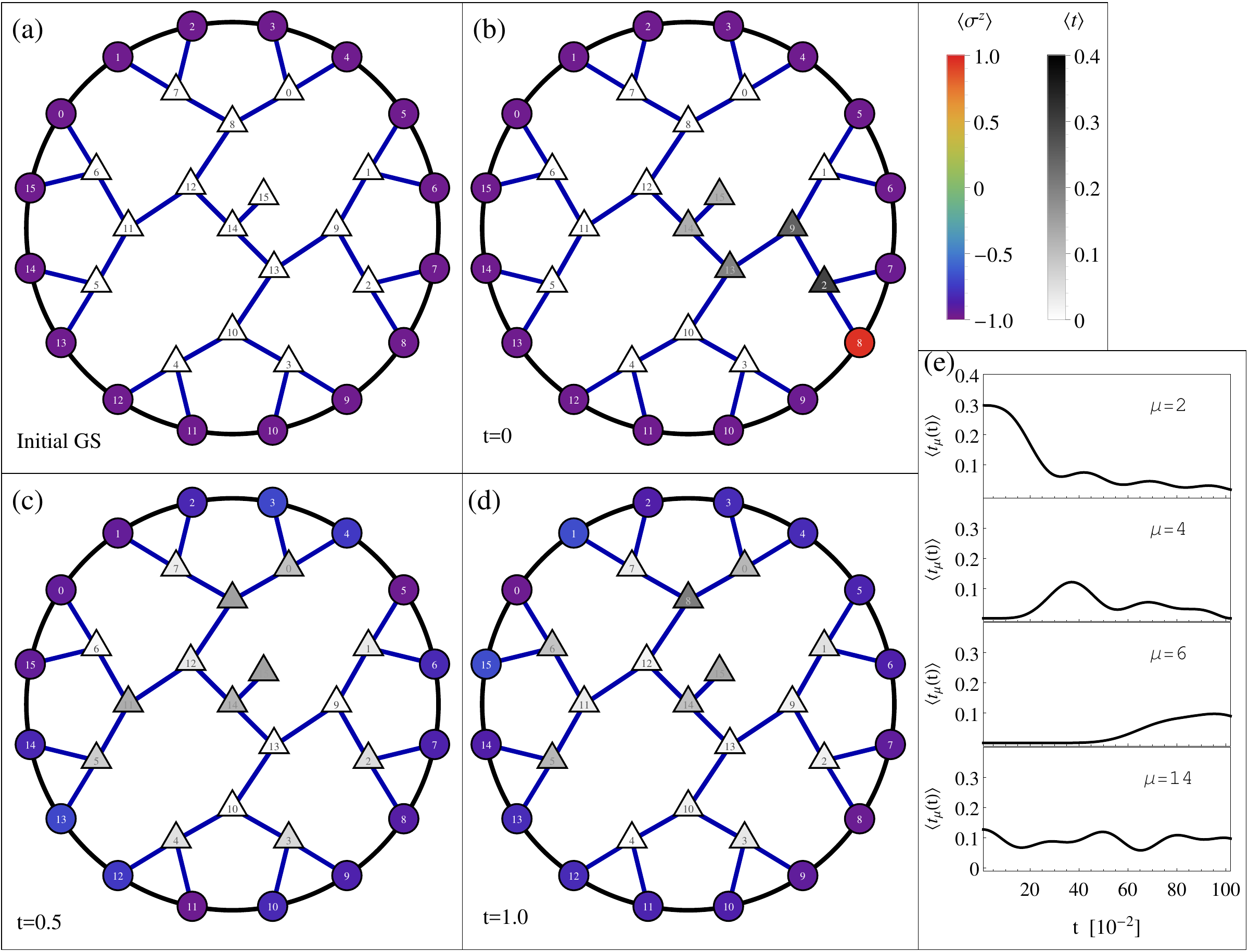}
	\caption{(color online) Combined plot of the physical $\langle\sigma^z_i(\text{t}) \rangle$ spins and bulk $\langle t_\mu(\text{t})\rangle$ qubits at several times during the time evolution. Interior triangles denote the bulk $t_\mu$ qubits where the isometries $W_\mu$ reside. The exterior circles are the physical boundary spins ${\sigma}_i$. The numbers are used to label boundary $\{i\}$ and bulk sites $\{\mu\}$. For clarity the disentanglers are not shown and the set of bulk sites are labeled in a cyclic order. (a) The initial ground state $|\Omega_0\rangle$ under the MERA circuit with uniform $\langle\sigma_i^z\rangle \approx -0.972$. The bulk qubits $t_\mu$ are strongly polarized toward zero with values in the $<10^{-3}$ regime. (b) Directly after acting with $\sigma^x_i$ at position $i=8$. With the exception of the excited spin, the rest of the physical spins $\sigma_i^z$ or rather their expectations values are little affected by the sudden excitation. But in the bulk degrees of freedom, there is a string of excited $\langle t_\mu\rangle >0$ qubits (located at $\mu=2,9,13,14,15$) emanating from the excitation at $i=8$. (c) After some time has elapsed $t=0.5$. The bulk excitations with $\langle t_\mu \rangle>0$ have now de-localized into the rest of the interior. (d) At the final time $t=1.0$. (e) Several time traces of $\langle t_\mu(\st)\rangle$ taken from selected bulk sites.} 
	\label{fig:MRing}
\end{figure*}

\begin{figure}
	\includegraphics[width=0.3\textwidth]{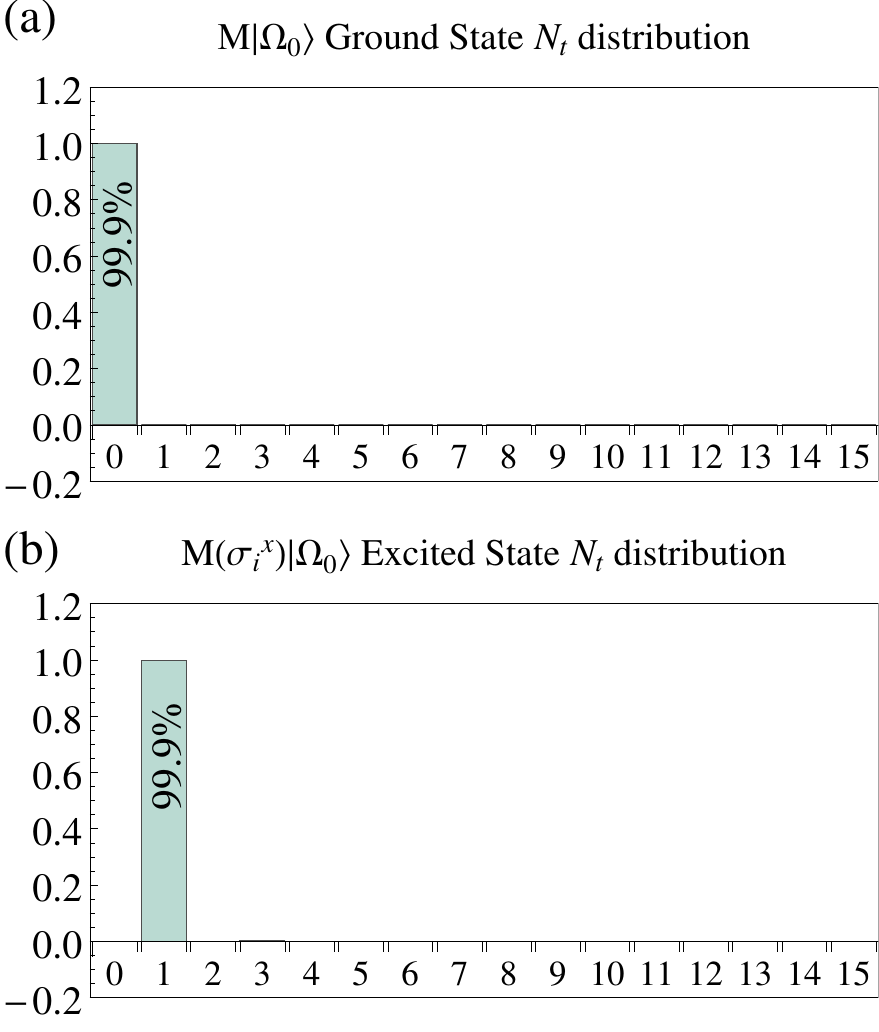}
	\caption{Spectral distribution of the total hologron number operator $N_t=\sum_\mu t_\mu$, for the transformed ground state $\MM|\Omega_0\rangle$ (a), and the transformed spin-flipped excited state $\MM \sigma^x_8 |\Omega_0\rangle$ (b). For $\MM|\Omega_0\rangle$ the spectral weight is strongly localized in the $N_t=0$ vacuum sector where the trivial product state $|0\rangle$ lies. For the excited state $\MM \sigma^x_8 |\Omega_0\rangle$, the spectral weight is now localized in the $N_t=1$ sector which is an $16$ dimensional subspace of $\mathcal{H}_t$.}\label{fig:weights}
\end{figure}

Next we apply the MERA gate $\MM$ to the time evolved excited state $|\Psi(\st)\rangle = \e^{-i 2\pi H \text{t}}\sigma^x_8 |\Omega_0\rangle$ and numerically compute the $\langle t_\mu \rangle$ observable in the bulk. This is numerically exact but technically tedious, and is performed by computing every coefficient of $\MM |\Psi(\st)\rangle$ in the $|\nn\rangle$ basis. Each coefficient computation corresponds to the contraction of a conventional coarse-graining MERA tensor network diagram with the input wavefunction $|\Psi(\st)\rangle$, but with the key difference that the isometries at each bulk node $\mu$ is substituted with either $W_{l\mu}$ or $W_{h\mu}$ depending on the configuration of $|\nn\rangle$. 

Shown in Fig.\ref{fig:MRing} are computations of the physical spin expectation value $\langle \sigma^z_i\rangle$ and the bulk qubit $\langle t_\mu \rangle$ expectation values at various times before and after the system is excited. For the ground state seen in Fig.\ref{fig:MRing}(a), the bulk qubits are very strongly polarized to $t_\mu =0$ indicating a high fidelity to the bulk product state vacuum $|0\rangle$. A spectral distribution of this state with respect to the total hologron number operator $N_t=\sum_{\mu} t_\mu$ is presented in Fig.\ref{fig:weights}(a). It shows the strong localization of the $\MM$ transformed exact ground state to the zero number sector which is inhabited by the trivial product hologron vacuum state $|0\rangle$. 
Hence, the large $h^z=3$ field has acted as a control parameter to achieve efficient disentanglement of the ground state by the MERA network, in the same spirit that the large-$N$ parameter in gauge-gravity duality acts to suppress gravitational quantum fluctuations. 
 
However immediately after the spin-flip is performed as seen in Fig.\ref{fig:MRing}(b), the bulk is altered by the appearance of a string of excited states with $\langle t_\mu\rangle>0$, emanating from the excited site at $i=8$. Its spectral distribution with respect to $N_t$ which is shown in Fig.\ref{fig:weights}(b), also exhibits a strong localization in $\mathcal{H}_t$. But now in the single hologron number sector $N_t=1$ which is a $L=16$ dimensional subspace of $\mathcal{H}_t$. After more time has progressed as shown in Fig.\ref{fig:MRing}(c,d), the bulk excitations de-localize from their original positions. More importantly, the spectral distribution remains strongly localized in the 1-hologron subspace over this time period. This suggests that the 1-hologron is a \emph{stable} holographic quasiparticle! 

In Fig.\ref{fig:MRing}(e) several time traces of $\langle t_\mu(\st)\rangle$ are taken from specific bulk sites. Note that the sites $\mu=2,4,6$ lie closest to the boundary and hence translate to physical degrees of freedom that are the most local. Because the bulk qubit at $\mu=2$ is situated closest to the physical excitation at $i=8$, it registers the greatest value of $\langle t_\mu \rangle$ at $\st=0$ but then relaxes over time. The bulk qubits at sites $\mu=4,6$ lie further away and require more time before the wave of $\langle t_\mu \rangle > 0$  arrives. Moreover, it takes the qubit at $\mu=6$ more time for this to occur than the qubit at $\mu=4$ because it is located further away from string of bulk excitations seen in Fig.\ref{fig:MRing}(b). This demonstrates that there is some degree of locality present in the dynamics of the bulk qubits because a finite amount of time is required for information to travel. It suggests a local description of the bulk dynamics that we shall discuss in the next section. Lastly, the qubit at $\mu=14$ that lies deep in the bulk experiences comparatively weaker fluctuations over time. This is not unexpected since its physical support is highly non-local and requires large scale fluctuations in the physical spins before it becomes comparable to the fluctuations seen at $\mu=2,4,6$. In addition, due to its large non-locality, the qubit at $\mu=14$ registers the spin-flip excitation instantaneously and lies along the string of bulk excitations of Fig.\ref{fig:MRing}(b).

\section{Effective Bulk Dynamics of Hologrons}\label{sec:HBulk}

In this section, we discuss in more detail the theoretical implications of the previous section's numerical results to make connection to ideas from AdS/CFT or gauge-gravity holography. We will continue to work within the paramagnetic phase with $h^z=3$, where there is a strong fidelity between the exact ground state $|\Omega_0\rangle$ and the MERA ansatz $\MM^\dagger|0\rangle$, such that we take them to be identical. We will mostly focus on the physical significance of the robust single hologron quasiparticle excitations that appear in this parameter regime.   

\subsection{Sourcing the bulk with boundary operators}\label{sec:sourcing}

\begin{figure}
	\includegraphics[width=0.475\textwidth]{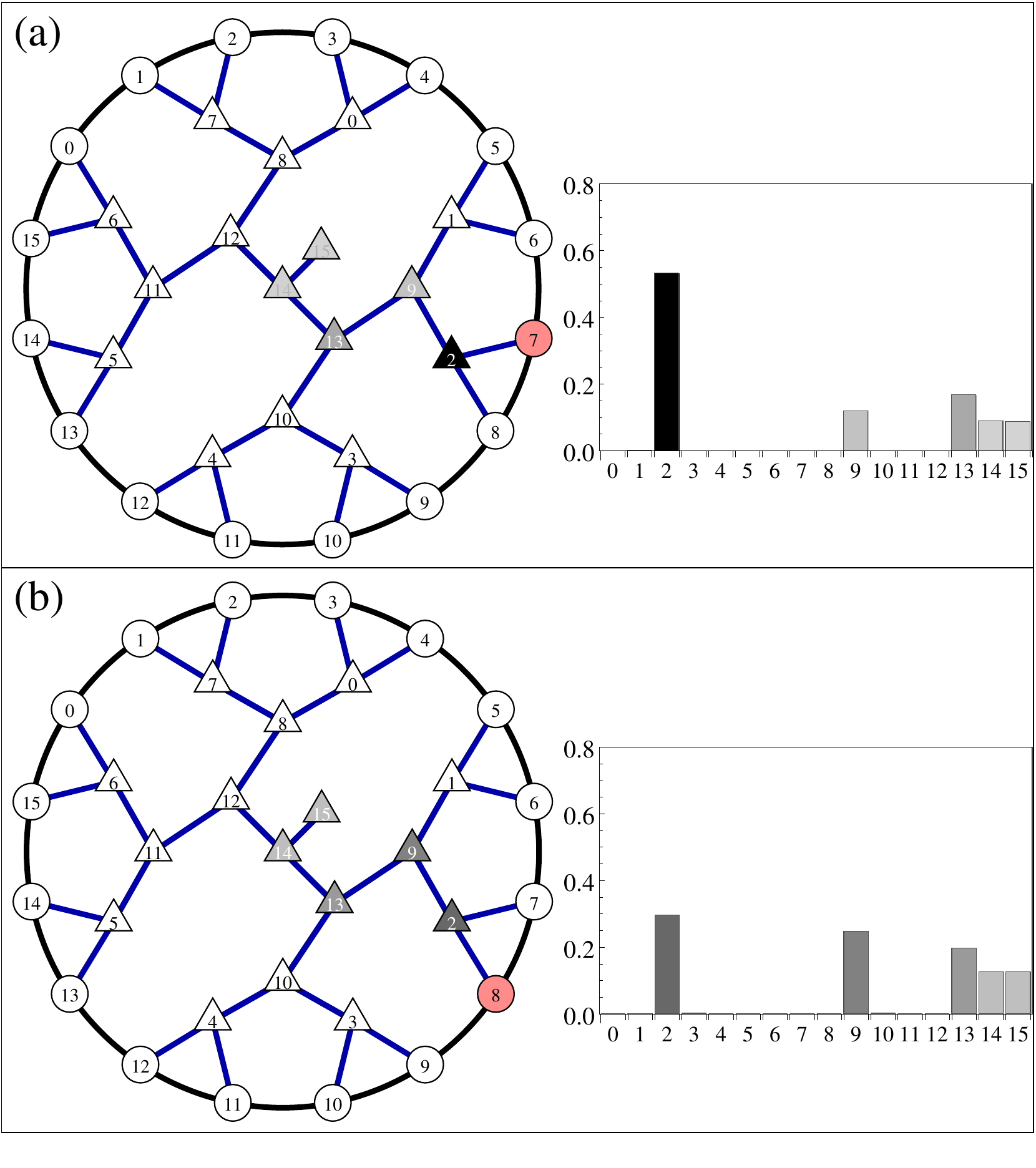}
	\caption{(color online) Plots of the 1-hologron bulk states $|J_i^x\rangle$ sourced by the spin-flip operators $\sigma^x_i$ acting on ground state $|\Omega_0\rangle$. Due to translational symmetry, only a pair of odd (a) and even (b) boundary sites are shown that are marked by the red disks. Note that the MERA network itself breaks the 1-site translational symmetry of the physical chain down to a 2-site translation. The histograms are probability density distributions of $|J_i^x\rangle$ within the bulk, with the bulk sites arranged along the binning axis. In either case, $|J_i^x\rangle$ is a string-like excitation in the bulk emanating from the physical boundary excited site, just as seen in Fig.\ref{fig:MRing}. }
	\label{fig:Jsources}
\end{figure}

Just like in gauge-gravity holography, physical operators such as ${\sigma}_i^\alpha$ acting on the physical ground state $|\Omega_0\rangle$ can be seen to create bulk excitations. This sourcing of hologron states in the bulk can be interpreted through the ``lifting" action\cite{evenbly2009algorithms} of $\MM$ on the operator in question. 

We define the MERA transformed spin-flip operator $\sigma^x_{i}$ by
\begin{align}
J_i^x := \MM \sigma_i^x \MM^\dagger
\label{eqn:Jix}\end{align}
which still satisfies $(J_i^x)^2= \mathbbm{1}$ and Pauli algebra with the analogous $J^{y,z}_i$ operators. Now from our numerical results of Sec.\ref{fig:weights}(b), the action of $\sigma_i^x$ on $|\Omega_0\rangle$, and equivalently of $J_i^x$ on $|0\rangle$ suggests the constrained expansion 
\begin{align}
J_i^x = \sum_{\mu \in \JJ^+(i)}\sum_{\alpha = x,y} (J^{x,\alpha}_{i,\mu}) \chi_\mu^\alpha
\label{eqn:Jix_exp}
\end{align}
where $\JJ^+(i)$ is the `causal cone' of site $i$ on the physical boundary as defined by the quantum circuit $\MM$. We have also used the notation $\chi^{x,y,z}_\mu$ to denote the Pauli gates $X_\mu,Y_\mu,Z_\mu$ acting on the bulk qubit $t_\mu$. This expansion is justified by the fact that the action on $|0\rangle$ as defined by 
\begin{align}
|J_i^x\rangle := J_i^x |0\rangle = \MM \sigma^x_i |\Omega_0\rangle 
\end{align}
is localized in the single hologron sector $\sum_{\mu} t_\mu = 1$, at least in the gapped phase with $h^z=3$. If there are other terms present on the RHS of (\ref{eqn:Jix_exp}), the condition that $\sum_{\mu} t_\mu |J_i^x\rangle = |J_i^x\rangle$ would be violated. Moreover since the space of operators $\{\chi_\mu^{\alpha}\}$ and the Clifford algebra it generates, act irreducibly on $|0\rangle$ to span $\mathcal{H}_t$, the expansion (\ref{eqn:Jix_exp}) is unique. Now there also exists a U(1) gauge redundancy in (\ref{eqn:Jix_exp}) generated locally in the bulk by $\chi^z_\mu$. In that we can rotate $\chi^{x,y}_\mu$ by $\e^{i \phi \chi^z_{\mu}}$ which is compensated by a transformation of the vector $(J_{i,\mu}^{x,x},J_{i,\mu}^{x,y})$. We then exploit this internal gauge symmetry of the MERA circuit to set, without loss of generality, $J_{i,\mu}^{x,y}=0$. We will make this gauge choice in the rest of paper. This leads to the simplified expansion,
\begin{align}
J_i^x = \sum_{\mu \in \JJ^+(i)} (J^{x}_{i,\mu}) \chi_\mu^x
\label{eqn:Jix_exp2}
\end{align}
where $J_{i,x}^x \equiv J_{i,\mu}^{x,x}$. 

In practice the coefficients $\{J_{i,\mu}^x\}$ have to be numerically determined which we have done. Shown in Fig.\ref{fig:Jsources} are plots of the 1-hologron excitations $|J_i^x\rangle$, as the appear in the bulk. Like previously observed in Fig.\ref{fig:MRing}(b), they have the character of string-like excitations that emanate from the boundary excitation site. In fact, because of the localization to the 1-hologron number sector ($N_t=1$) the amplitude squared $|J_{i,\mu}^x|^2$ for fixed $i$ is identical to $\langle t_\mu(0)\rangle$ of Fig.\ref{fig:MRing}(b). Also as time progresses, we know from the real-time dynamics data, that the string excitation will subsequently diffuse in the bulk. In Section \ref{sec:ME} we determine the specific bulk 1-hologron Hamiltonian which drives this diffusion. 

A physical interpretation of the state $|J_i^x\rangle$ is that, it is the result of sourcing hologrons using the `magnon' creation operator ${\sigma}^x_i$ acting in the spin-ordered phase $\langle \sigma_i^z\rangle \neq 0$. Thus $|J_i^x\rangle$ is the holographic dual to an Ising magnon mode, which because of the Hamiltonian, will dynamically evolve over time. It must be emphasized that the excited state ${\sigma}^x_{i} |\Omega_0\rangle$ is still a complicated many-body state. However the MERA transform acts to simplify the ground state $|\Omega_0\rangle$ into a trivial product state $|0\rangle$ at the cost of making the local excitation operator $\sigma_i^x$ non-local in one dimension higher. This operator in the bulk is of course $J_i^x$ of equation (\ref{eqn:Jix_exp2}).

\begin{figure}
	\includegraphics[width=0.3\textwidth]{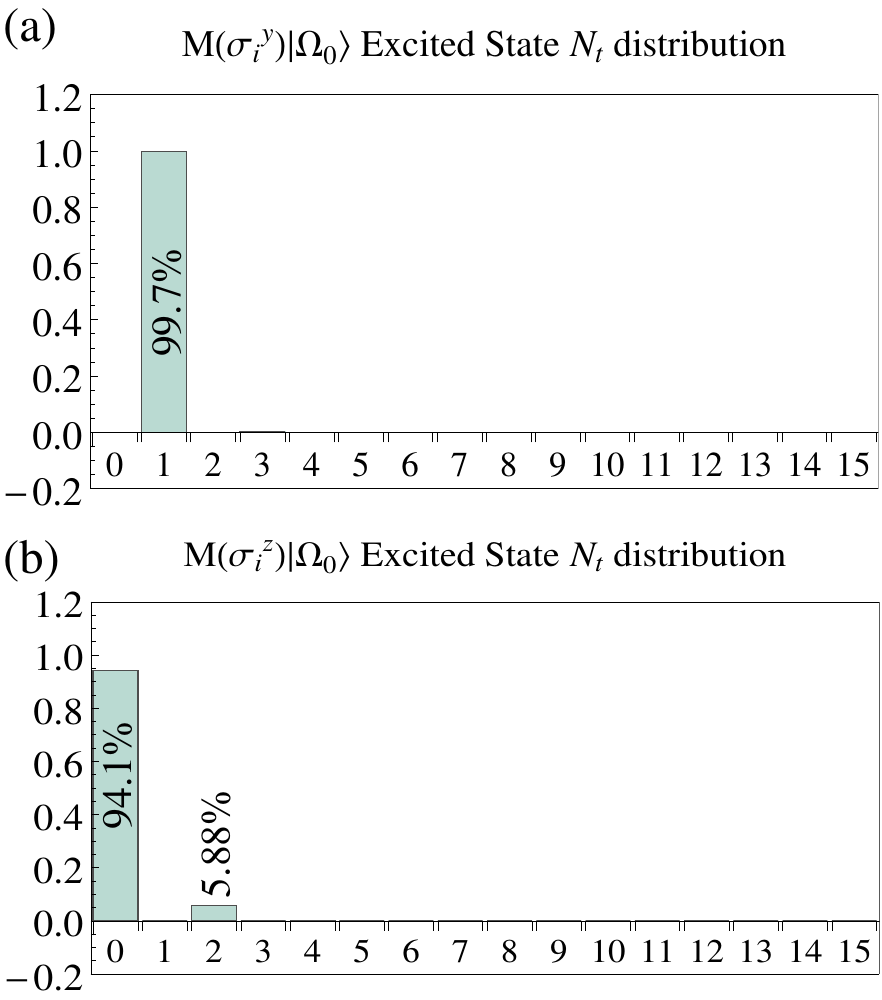}
	\caption{Spectral distribution of the total hologron number operator $N_t=\sum_\mu t_\mu$, for the transformed spin-flipped excited state $\MM \sigma^y_8 |\Omega_0\rangle$ (a) and the spin-flipped excited state $\MM \sigma^z_8 |\Omega_0\rangle$ (b). For $\MM \sigma^y_8|\Omega_0\rangle$ the spectral weight remains strongly localized in the $N_t=1$ while for the state $\MM \sigma^z_8 |\Omega_0\rangle$, the spectral weight is now distributed over the $N_t=0,2$ sectors.}\label{fig:weights4}
\end{figure}

Analogous operator expansions hold for the $J_i^y, J_i^z$ boundary sourced operators which remain strongly localized in a few sectors in $\{\chi^\alpha_\mu\}$ operator space, as is shown in Fig.\ref{fig:weights4}. 
They are given as
\begin{align}
J_i^y &= \sum_{\mu \in \JJ^+(i)} (J^{y}_{i,\mu}) \chi_\mu^y \\
J_i^z &= \sum_{\mu \in \JJ^+(i)} (J^{z,z}_{i,\mu}) \chi_\mu^z
+\sum_{\mu,\nu \in \JJ^+(i)} \; \sum_{\alpha,\beta =x,y} (J^{z,\alpha\beta}_{i,\mu\nu}) \chi_\mu^\alpha \chi_\nu^\beta 
\end{align}
Like the $J_i^x$ operator, the expansion of  $J_i^y$ involves only superpositions of single $\chi^{x,y}_\mu$ operators. But the expansion of $J_i^z$ involves superpositions of single $\chi^z_\mu$ and double spin-flip operators $\chi^{x,y}_\mu \chi^{x,y}_\nu$. Moreover, from the expansions of $J_i^x$ and $J_i^y$ it is clear that the parity of the hologron number $N_t$ and the parity of the number of spin-flips ($\sigma^{x,y}_i$) agree. This reflects the $\mathbb{Z}_2$ symmetry of the transverse field Ising model, which becomes a number parity symmetry of the Jordan-Wigner fermions. Since $\sigma_i^{x,y,z}$ for all $i$ forms the full set of generators of the boundary observables, the full dictionary between boundary and bulk observables is furnished by this isomorphisms of Clifford algebras given by these operator expansions! 

\subsection{Physical States of the Hologrons}\label{sec:phys_hol}

\begin{figure*}
	\includegraphics[width=0.95\textwidth]{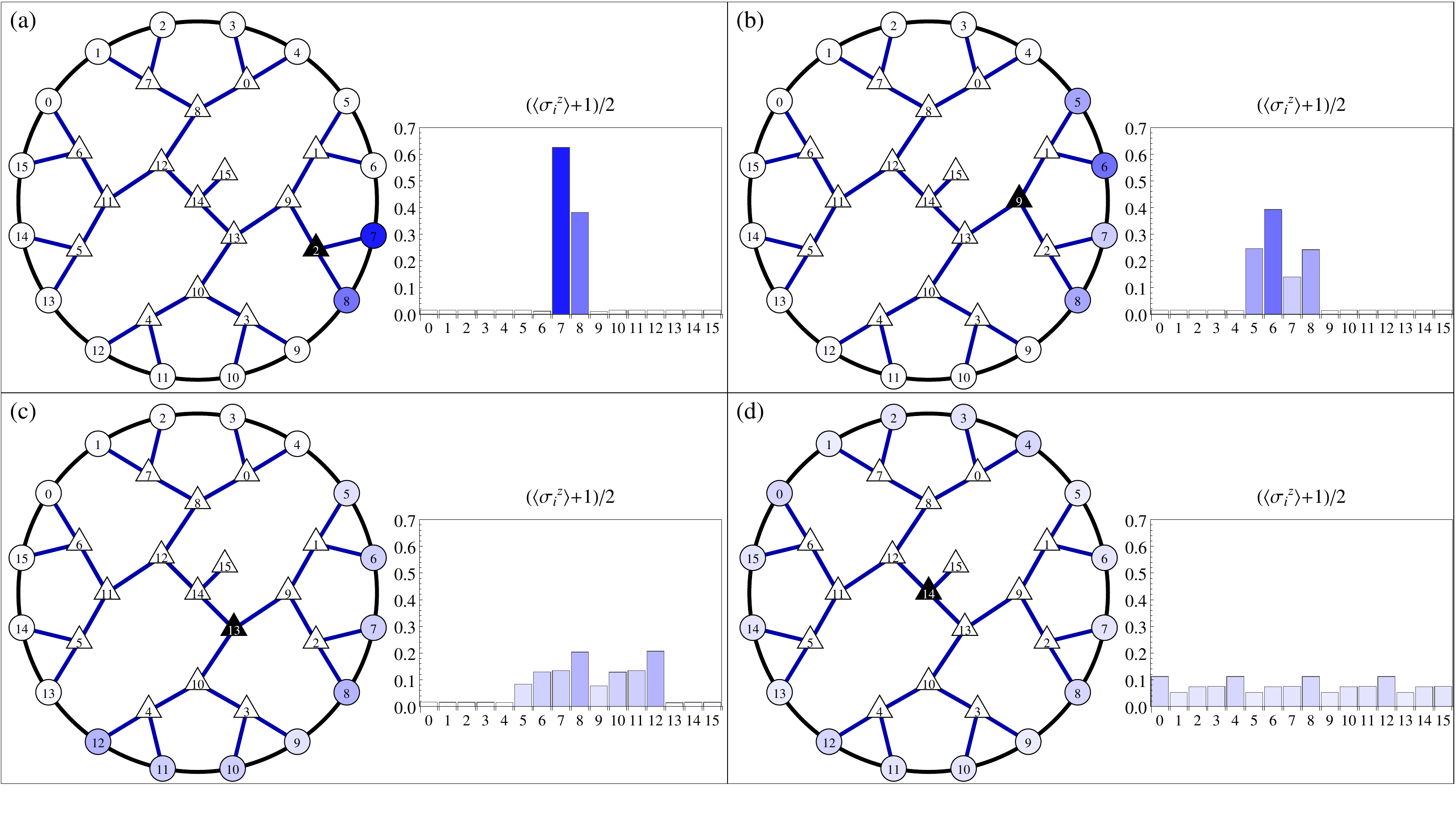}
	\caption{(color online) The boundary states of local bulk hologrons $\MM^\dagger \chi^x_\mu|0\rangle \in \mathcal{H}_s$. The dark triangles mark the location $\mu$ of the hologron in the bulk. In the paramagnetic phase $h^z=3$, the physical observable $\langle \sigma^z_i\rangle$ is strongly polarized towards $-1$. So shown in the bar chart are the observables $(\langle \sigma_i^z\rangle +1)/2$ as a function of site $i$, and are strongly polarized towards 0 in the ground state. Due to translational symmetry only 4 chosen hologron states(a,b,c,d) are shown, with increasing circuit depth $\tau$. But there is a slight left-right asymmetry from the MERA tensors network and is not a feature of the physical model. As was mentioned in Section \ref{sec:real_dynamics}, the hologrons  closer (further) to the boundary are more localized (non-local) in physical space. Again, this can be understood purely from the structure of their causal cones. For the two deepest hologrons located at $\mu=14,15$, their physical influence extends to the entire boundary. Otherwise, when the physical extent of a local bulk hologron excitation is bounded on the boundary, its physical state resembles a `stretched magnon' waveform or wavelet.}
	\label{fig:grav_wave}
\end{figure*}

Next we take the dual viewpoint and consider the sorts of physical states that are produced by single bulk hologrons. In this instance, the boundary wavefunctions $\MM^\dagger \chi_\mu^x |0\rangle \in \mathcal{H}_s$ were computed from exciting the trivial state $|0\rangle$ and then applying the inverse MERA transform $\MM^\dagger$. Technically, this is achieved by replacing at $\mu$ the low isometry $W_{l\mu}$ in the ground state ansatz network with a high isometry $W_{h\mu}$ and then producing an entire wavefunction in  $\mathcal{H}_s$ with the modified tensor network. 

Shown in Fig.\ref{fig:grav_wave} are plots of the local observables
$s_i \equiv (\langle \sigma_i^x\rangle +1)/2$ computed with $\MM^\dagger \chi_\mu^x |0\rangle$ for all $\mu$. It shows that the local bulk hologron states produce non-local changes to the otherwise ordered ground state $|\Omega_0\rangle$; their physical range depending on how deep in the bulk these hologrons are. Physically, the 1-hologron states resemble non-local `stretched magnon' excitations on the boundary. A sort of holographic analogue of the Ising magnon $\sigma_i^x |\Omega_0\rangle$, which is the physical content of equations (\ref{eqn:Jix_exp}) and (\ref{eqn:Jix_exp2}). Hence the complimentary viewpoint that local bulk degrees of freedom are non-local on the boundary, is confirmed from these single hologron states. 

\subsection{Bulk-Boundary Correspondence}

To complete the ``holographic" correspondence between the physical (boundary) and ancillae (bulk) degrees of freedom, we proceed by defining the partition (generating) function on the boundary with respect to the Heisenberg operator $\sigma^x_i(\st):= \e^{i H \st} \sigma_i^x \e^{-i H \st}$ as
\begin{align}
\mathcal{Z}_\text{bnd}[h_i^x] :=\langle \Omega_0 | T \exp\left( -i \int_{\st_i}^{\st_f} \mathrm{dt}\sum_i h_i^x(\mathrm{t})\sigma_i^x(\mathrm{t}) \right) |\Omega_0\rangle
\label{eqn:Zbnd}
\end{align}
such that the time-ordered correlators are obtained from differentiation
\begin{align}
\langle \Omega_0 | T \sigma_{i_1}^x(\st_1) \sigma_{i_2}^x(\st_2) |\Omega_0 \rangle = i^2\left.\frac{\delta^2 \mathcal{Z}_\text{bnd}[h_i^x]}{\delta h^x_{i_1}(\st_1) \delta h^x_{i_2}(\st_2)}\right|_{h_i^x=0}
\end{align}
and more generally
\begin{align}
\langle \Omega_0| T \prod_{n=1}^N \sigma_{i_n}^x(\st_{n})|\Omega_0\rangle
= i^N\left(\prod_{n=1}^N \frac{\delta}{\delta h_{i_n}^x(\st_{n})} \right) \mathcal{Z}_\text{bnd}[0].
\end{align}
A few comments are in order for readers more used to quantum field theory. Here we have used the canonical formulation with Dyson time-ordering in defining the correlation functions as opposed to the more popular functional integral formulation. Our reasons have primarily to do with simplicity. Nevertheless it is well known that formulating the partition function of the transverse field Ising model leads to the 2D Ising model.\cite{kogut1979introduction} It would be worthwhile in the future to flesh out the detailed ramifications of the MERA quantum circuit from the 2D Ising model perspective. We also note that our formulation of spin-1/2 correlators and their associated equation of motions can be cast into the Schwinger equations of motion framework developed in Ref.[\onlinecite{Shastry2014164}] for the more general case of Hubbard operators. 

Secondly, we have defined the ``interaction picture" with respect to the full Hamiltonian \[H=-\sum_{i}\sigma_i^x \sigma_{i+1}^x + h^z\sigma_i^z\] since there are no ``free Gaussian terms" in the Hamiltonian from the point of view of spins. Thus there is no need to divide by ``vacuum bubbles" and for the notion of ``in" and ``out" states at asymptotic infinity. The partition function $\mathcal{Z}_\text{bnd}$ should then really be identified with the  ground state to ground state quantum amplitude  in the presence of sources, in real time. 

Thirdly, the addition of the source field is equivalent to adding a time-dependent source term to the Hamiltonian such that $H\rightarrow H + \sum_i h^x_i(\st)\sigma^x_i$ where $h_i^x$ can be seen as a local longitudinal field. Moreover since the transverse field Ising model is exactly integrable by the Jordon-Wigner(JW) transform, $\mathcal{Z}_\text{bnd}[h_i^x]$ can be computed order by order in $h_i^x$ by Gaussian integration of the JW fermions. 

\begin{figure}
	\includegraphics[width=4.5cm]{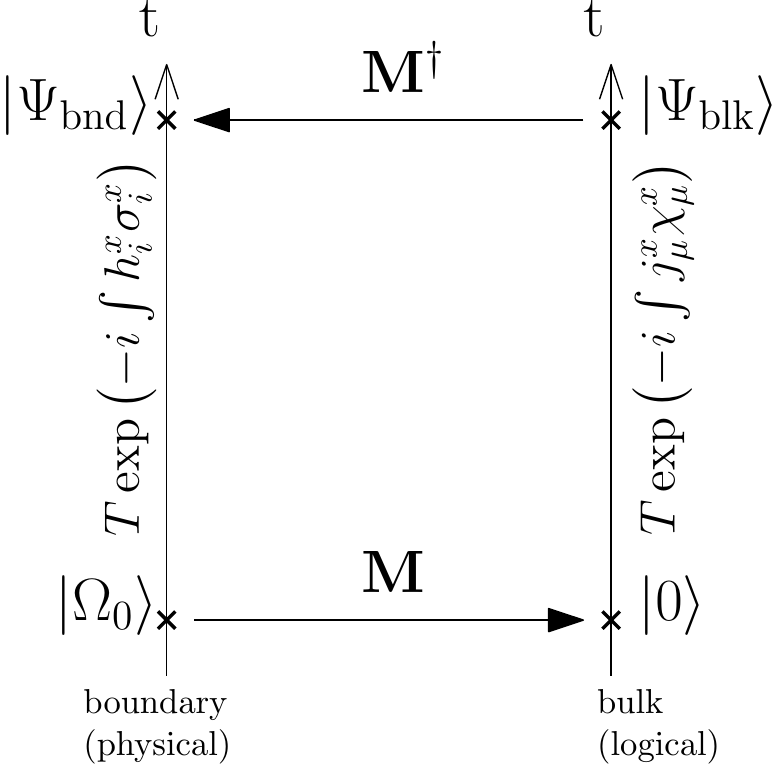}
	\caption{A schematic of the bulk-boundary correspondence which is just a unitary isomorphism between physical $\mathcal{H}_s$ and logical or ancillae $\mathcal{H}_t$ Hilbert spaces. The vertical axis denotes real-time $\st$ direction and the unitary maps $T\exp\left(-i \int h_i^x \sigma^x_i \right)$ and $T\exp\left(-i \int j_\mu^x \chi^x_\mu \right)$ act on $|\Omega_0\rangle$ and $|0\rangle$ respectively. The respective partition functions or quantum amplitudes are given by the overlaps $\mathcal{Z}_\text{bnd}=\langle \Psi_\text{bnd}|\Omega_0\rangle$ and $\mathcal{Z}_\text{blk} = \langle \Psi_\text{blk}|0\rangle$. The holographic equivalence is the statement that $\mathcal{Z}_\text{bnd}=\mathcal{Z}_\text{blk}$ which derives from $|\Omega_0\rangle = \MM^\dagger |0\rangle$ and the relation (\ref{eqn:sources}) between the source fields.}
	\label{fig:square}	
\end{figure}

Moving on, we next apply the MERA quantum circuit to the 1D chain and take as an equality that the ansatz and ground state agree, $\MM^\dagger |0\rangle = |\Omega_0\rangle$. Then the partition function transforms as
\begin{align}
\mathcal{Z}_\text{bnd}[h^x_i]&=\langle \Omega_0 | T \exp\left( -i \int_{\st_i}^{\st_f} \mathrm{dt}\sum_i h_i^x(\mathrm{t})\sigma_i^x(\mathrm{t}) \right) |\Omega_0\rangle \nonumber \\
&= \langle 0 | T\exp\left(
-i \int_{\st_i}^{\st_f} \mathrm{dt} \sum_{\mu} j_\mu^x(\mathrm{t}) \chi^x_{\mu}(\mathrm{t})
\right)|0\rangle \nonumber \\
&= \mathcal{Z}_\text{blk}[j_\mu^x]
\end{align}
where from (\ref{eqn:Jix_exp2}), the source field $h_i^x$ on the physical Hilbert space $\mathcal{H}_s$ is related to the source field $j_\mu^x$ in the bulk Hilbert space $\mathcal{H}_t$ by 
\begin{align}
j_\mu^x (\mathrm{t}) := \sum_i (J_{i,\mu}^{x}) h_i^x(\mathrm{t}).
\label{eqn:sources}
\end{align}
That is $j_\mu^x$ satisfies the ``boundary condition" imposed by $h_i^x$ and $(J^x_{i,\mu})$ the linear transformation (kernel or matrix) that relates the two. These coefficients were computed with the MERA tensors and are basically the numerical results shown in Fig.\ref{fig:Jsources}. The statement that boundary and bulk partition functions agree $\mathcal{Z}_\text{bnd}[h_i^x] = \mathcal{Z}_\text{blk}[j^x_\mu]$ for suitable matched source terms is the \emph{holographic duality}! This relationship is visually expressed in Fig.\ref{fig:square}. 

On the surface, this duality trivially appears as a unitary or canonical transform implemented by $\MM$. However from a correlations and entanglement perspective the vacuum states $|\Omega_0\rangle, |0\rangle$ used to evaluate the respective partition functions are vastly different. In that, the boundary state $|\Omega_0\rangle$ contains quantum ground state correlations, whilst the bulk state $|0\rangle$ is an exact tensor product and is completely trivial in terms of the logical qubits $t_\mu$ in the bulk. Thus the duality implemented by $\MM$ is a `\emph{correlated-uncorrelated}' duality between the physical (boundary) $s_i$ and the logical (bulk) $t_\mu$ degrees of freedom. It is very tempting to label this a ``strong-weak" duality as in commonly done in the AdS/CFT literature, however we must remind ourselves that the boundary Hamiltonian is really a theory of free JW fermions, but at a finite occupation which gives rise to the non-trivial ground state correlations.
Moreover, here the bulk remains fully quantum mechanical as opposed to the AdS/CFT correspondence where the bulk is approximately classical in the large-$N$ limit. Even though the large $h^z$ plays a role that is similar spirit to the large-$N$, we should emphasize that the similarities are only superficial, mainly because there is no `dynamical' gravity in the MERA network and the model is in the non-CFT limit.

Although it is simple to express $\mathcal{Z}_\text{blk}[j_\mu^x]$, the complexities of the dynamics of the bulk operators $\chi^\alpha_\mu(\st)$ are encoded in their equations of motion
\begin{align}
\partial_\st \chi^\alpha_\mu(\st) &= i[H_\MM, \chi_\mu^\alpha(\st)] 
\end{align} 
with the bulk Hamiltonian $H_\MM = \MM H \MM^\dagger$. In practice this is done through the study of its correlators or Green's functions. In general the multi-point ($n>2$) Green's function satisfies complicated equations of motions because of the complicated matrix elements of $H_\MM$. They have to be studied in the different excitation sectors at a time. Fortunately we can organize the correlators according to the number of $\sigma_i^x$ insertions which corresponds to the number of different hologron number $N_t = \sum_{\mu} t_\mu$ sectors. In the next subsection, we will discuss our numerical results from just the $N_t=1$ single hologron sector $\mathcal{H}_t^{(1)}$.

\subsection{The bulk Hamiltonian $H_M$ and the single hologron sector $\mathcal{H}_t^{(1)}$}\label{sec:ME}

\begin{figure}
	\includegraphics[width=0.475\textwidth]{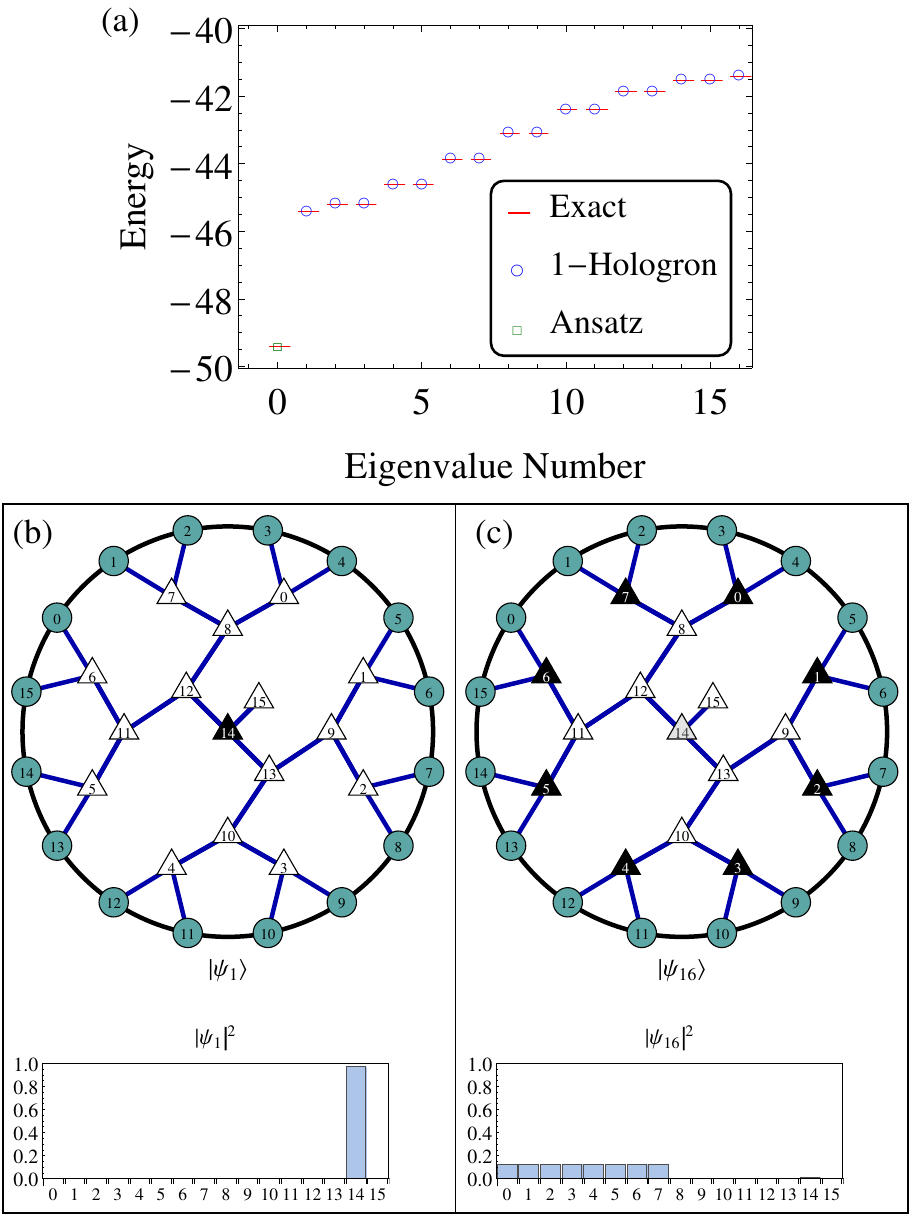}
	\caption{(color online) The spectrum of $H^{(1)}_\MM$ which is $H_\MM$ truncated to the 1-hologron subspace $\mathcal{H}^{(1)}_t$. The model remains in integrable paramagnetic phase with $h^z=3$, and the overall agreement between the exact and effective is excellent. (a) The comparison between the exact low energy spectrum of $H$ and $H^{(1)}_M$, as well the variational ansatz energy $\langle 0 | \MM H \MM^\dagger |0\rangle$. (b), (c) Wavefunctions of the first $|\psi_1\rangle$ and $16^\text{th}$ $|\psi_{16}\rangle$, excited eigenstates of $H_\MM$ computed in the 1-hologron sector. The histograms display the probability densities at each bulk site. $|\psi_1\rangle$ is strongly held at the bulk center while $|\psi_{16}\rangle$ is evenly distributed near boundary.}
	\label{fig:energy1}
\end{figure}

The matrix elements of the bulk Hamiltonian $H_M$ can be straightforwardly computed by essentially the same methods as used to compute observables with MERA tensors.\cite{evenbly2009algorithms}  The technical details are outlined in Appendix \ref{app:MethodHam}. 

The results of Section \ref{sec:real_dynamics}, specifically Fig.\ref{fig:weights}(b), suggest that the low energy structure is described by the single hologron subspace which is precisely defined as \[\mathcal{H}^{(1)}_t := \text{Span}\{|1_\mu\rangle\equiv\chi^x_\mu|0\rangle, \mu \in \text{bulk}\}.\]Furthermore we define the 1-hologron Hamiltonian by
\begin{align}
H^{(1)}_\MM = \sum_{\mu,\nu\in \text{bulk}} \langle 1_\mu | H_\MM |1_\nu \rangle |1_\mu\rangle \langle 1_\nu| 
\end{align}
and compare the exact energy spectrum to the spectrum of the bulk Hamiltonian in the low energy effective 1-hologron subspace. Shown in Fig.\ref{fig:energy1}(a) is a comparison between the exact low energy spectrum of $H$ and the energy spectrum of the effective bulk Hamiltonian $H_M^{(1)}$, obtained by restricting $H_\MM$ to $\mathcal{H}_t^{(1)}$. A very good agreement is obtained between the exact and effective spectra, which is another indication of the long lived stability of the single hologron as a quasiparticle. However the higher excited states involving larger $N_t>1$ sectors are no longer reliably described by a `gas' of independent hologrons. Nevertheless, we can expect that in the dilute limit or small $N_t$, the accuracy of an independent hologron approximation will improve with increasing system size $L$, mainly due to the short range correlations of the massive paramagnetic phase. Also shown in Fig.\ref{fig:energy1}(b,c) are the lowest and highest energy eigenstates of $H_\MM^{(1)}$ as they appear in the bulk. The first excited state $|\psi_1\rangle$ is strongly localized at the center of bulk. Whereas $|\psi_{16}\rangle$, the highest excitation is evenly distributed along the bulk sites closest to the boundary, which are also physically the most UV in physical character. Generically, the lower energy the excitation, the more IR it is and its bulk state wavefunction remains closer to the center, and vice versa. This generally agrees with the intuition that IR (UV) degrees of freedom are lower (higher) in physical energy.

\begin{figure*}
	\includegraphics[width=0.9\textwidth]{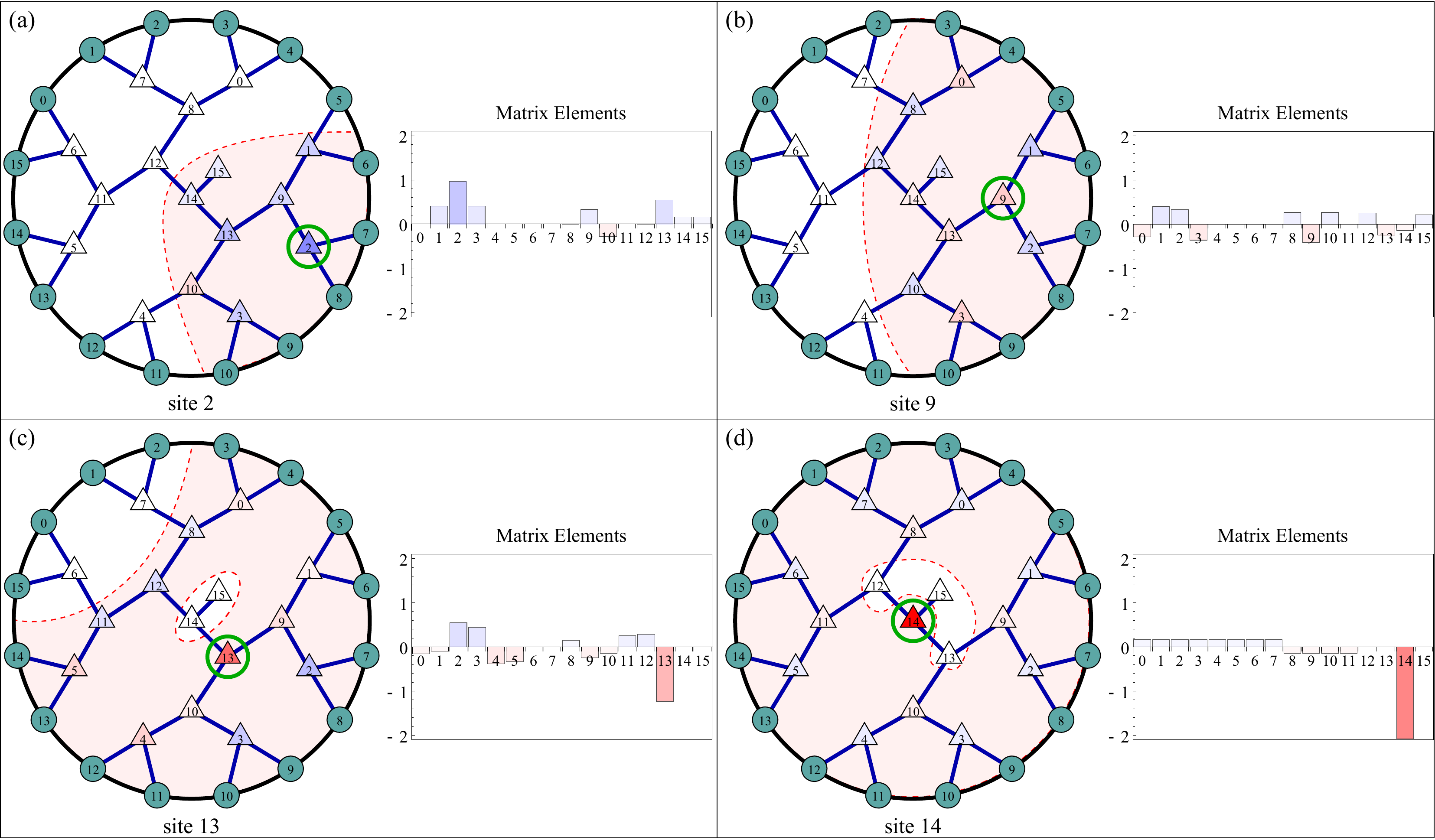}
	\caption{(color online) The matrix elements of $H^{(1)}_\MM$ interpreted as a hopping Hamiltonian in the bulk. Recall also that disentanglers are present but not drawn here. All matrix elements are all real-valued because the $H$ and the MERA tensors are real-valued. For clarity, only the traceless part of $H^{(1)}_\MM$ is plotted. The solid (green) circle marks the bulk site from which the hologron hops from. The triangles are colored according to the value of the relevant matrix element, and the triangle within the (green) circle corresponds to the on-site energy matrix element. The bar charts display the same data in quantitative values. For example in (d), the large negative value at site 14 indicates a deep potential well for the hologron. The shaded region bounded by the dashed (red) lines denotes the bulk sites accessible by hopping from the circle. The range of hopping is limited by both the causal cone structure (kinematics) and the specific values of the Hamiltonian (dynamics). For example the `forbidden pockets' around sites 12,13,14 and 15 in (c,d) are dynamical in origin.}
	\label{fig:ME}
\end{figure*}

Next, we examine the actual matrix elements of $H^{(1)}_\MM$, which can  be interpreted as a hopping Hamiltonian for a single hologron. Shown in Fig.\ref{fig:ME} is data from these matrix elements involving bulk sites of varying depths within the bulk. Most striking is Fig.\ref{fig:ME}(a) which exhibits the matrix elements from a near boundary (UV) bulk site. It displays the finite range of hopping into the bulk which can be traced back to the limited extent of the causal cones of the bond operators $H_i$ which lie under the influence of site $\mu=2$. This is a clear demonstration of the kinematical constraints on the 2 dimensional motion of the hologron quasiparticle. That although the bulk Hamiltonian $H_M$ is now non-local, the arrangement of the MERA network limits the degree of non-locality in the bulk. This is the main reason for the bulk locality that was alluded to in interpreting the data of Fig.\ref{fig:MRing}(e). Furthermore with increasing depth, the range of hopping increases. But there are sometimes pockets of forbidden regions, like in Fig.\ref{fig:ME}(c,d) which we regard as having to do with dynamical constraints imposed by the specific values of $H^{(1)}_\MM$. In the CFT limit of the holographic correspondence there are no specific limitations to bulk dynamics, save for conformally invariant data, such that the bulk is entirely described by kinematics alone, which is the observation of Ref.\onlinecite{czech2015tensor}. Lastly, although the hoppings from the deepest sites $\mu=14,15$ are fully non-local, the large negative on-site values [See Fig.\ref{fig:ME}(d)], indicate a deep potential well for the hologron and hence limit the effects of this non-locality by suppressing the quantum tunneling from the center. The deep well also explains the wavefunction localization of $|\psi_1\rangle$ of Fig.\ref{fig:energy1}(b). It is tempting to speculate that in the high-density ($N_t \gg 1$) thermal states, a deep central well in the bulk will aid in the formation of a `black hole' by condensation of hologrons.  

\subsection{Dynamics in the 1-hologron sector}

In this section, we study the real-time dynamics implied by $H_M^{(1)}$ as a low energy effective Hamiltonian. Again, we limit ourselves to the first $L$ eigenstates of the theory, or equivalently the 1-hologron subspace $\mathcal{H}^{(1)}_\MM$ which by (\ref{eqn:Jix_exp2}) is spanned by the single spin-flipped states $\sigma^x_i |\Omega_0\rangle$. The single hologron time-ordered correlator is defined to be 
\begin{align}
\mathrm{G}_{\mu\nu}(\st) := \langle 0 | T \chi^x_{\mu}(\st) \chi^x_\nu (0)|0\rangle 
\end{align}
and is related to the physical time-ordered correlation function 
\begin{align}
G_{ij}(\st) := \langle \Omega_0 | T \sigma^x_i (\st)\sigma^x_j(0)|\Omega_0\rangle 
\end{align}
through the time independent relation
\begin{align}
G_{ij}(\st) = \sum_{\mu,\nu} J_{i,\mu}^x \mathrm{G}_{\mu\nu}(\st) J_{j,\nu}^x.
\label{eqn:Gblk_Gbnd}
\end{align}
As usual, these can be obtained in principle through functional derivation of their respective partition functions. The causal (Feynman time-ordered) Green's function can be expressed in terms of the familiar retarded $\GG^R$ and advanced $\GG^A$ Green's functions,
\begin{align}
\GG_{\mu\nu}(\st) &= \;\;\;
\Theta(\st) \langle 0 | \chi^x_{\mu}(\st) \chi^x_\nu(0)|0\rangle \nonumber \\ &\;+
\Theta(-\st) \langle 0 | \chi^x_\nu(0) \chi^x_{\mu}(\st)|0\rangle \nonumber \\
&= \GG^R_{\mu\nu}(\st) + \GG^A_{\mu\nu} (\st).
\end{align}
It is convenient to associate the traceless part of $H^{(1)}_\MM$ with the Hamiltonian of the excitation 
\begin{align}
h^{(1)}_\MM = H^{(1)}_\MM - E_1 \mathbbm{1},\quad
E_1 = \left(\tfrac{1}{L}\right)\text{Tr}H^{(1)}_\MM.
\end{align}
The matrix elements of $h^{(1)}_\MM$ were shown in Fig.\ref{fig:ME} which demonstrate their varying degree of locality in the bulk. The difference in energy $(E_1-E_0)$ can be associated with the ``rest mass" of the single hologron excitation. An alternative and equally valid convention is to choose $E_1$ as the minimum of energy of $H^{(1)}_\MM$. The 1-hologron retarded/advanced Green's function then has the spectral expression
\begin{align}
\GG^{R}_{\mu\nu}(\st) &= \int_{-\infty}^\infty \mathrm{d}\omega \;\GG^{R}_{\mu\nu} (\omega) \;\e^{-i2\pi\omega \st} \\
\GG^{R}_{\mu\nu}(\omega) &= \left[\frac{i}{\omega- h^{(1)}_\MM - (E_1 -E_0) + i0^+}\right]_{\mu\nu}\\
\GG^A_{\mu\nu}(\omega) &= \GG^R_{\mu\nu}(-\omega)
\end{align}
where we have used the fact that $h^{(1)}_\MM$ is real and symmetric to derive $\GG^A$. This gives the form of the time-ordered correlator as 
\begin{align}
\GG_{\mu\nu} (\st) &= \int_{-\infty}^\infty \mathrm{d}\omega \;\GG_{\mu\nu} (\omega) \;\e^{-i2\pi\omega \st}  \\
\GG_{\mu\nu} (\omega) &= \GG_{\mu\nu}^R(\omega) +
\GG_{\mu\nu}^R(-\omega)  
\end{align} 
In practice, it is more straightforward to diagonalize $h^{(1)}_\MM$ and its eigenstates $\{|\varepsilon_n\rangle\}_n$ and compute $\GG^R_{\mu\nu}$ through
\begin{align}
\GG^R_{\mu\nu} (\st) &= \Theta(\st)\e^{-i2\pi(E_1-E_0)\st} \sum_{n} 
\langle \mu |\varepsilon_n\rangle \langle \varepsilon_n | \nu\rangle\; \e^{-i 2\pi{\varepsilon_n } \st }
\end{align}
where we associate the $n$ independent phase factor as being the oscillation due to the ``mass" of the excitation. This leads to the time-ordered correlation function in real time as
\begin{align}
\GG_{\mu\nu}(\st) &= \sum_n \langle \mu |\varepsilon_n \rangle \langle \varepsilon_n |\nu\rangle 
\e^{-i 2\pi(\varepsilon_n + E_1-E_0)|\st|} \nonumber \\
& = \langle \mu | \e^{-i 2\pi(h^{(1)}_\MM +E_1-E_0)|\st|} |\nu\rangle 
\end{align} 
In our numerics we will take $\st>0$ for simplicity and choose only the retarded branch.

\begin{figure}
	\includegraphics[width=0.475\textwidth]{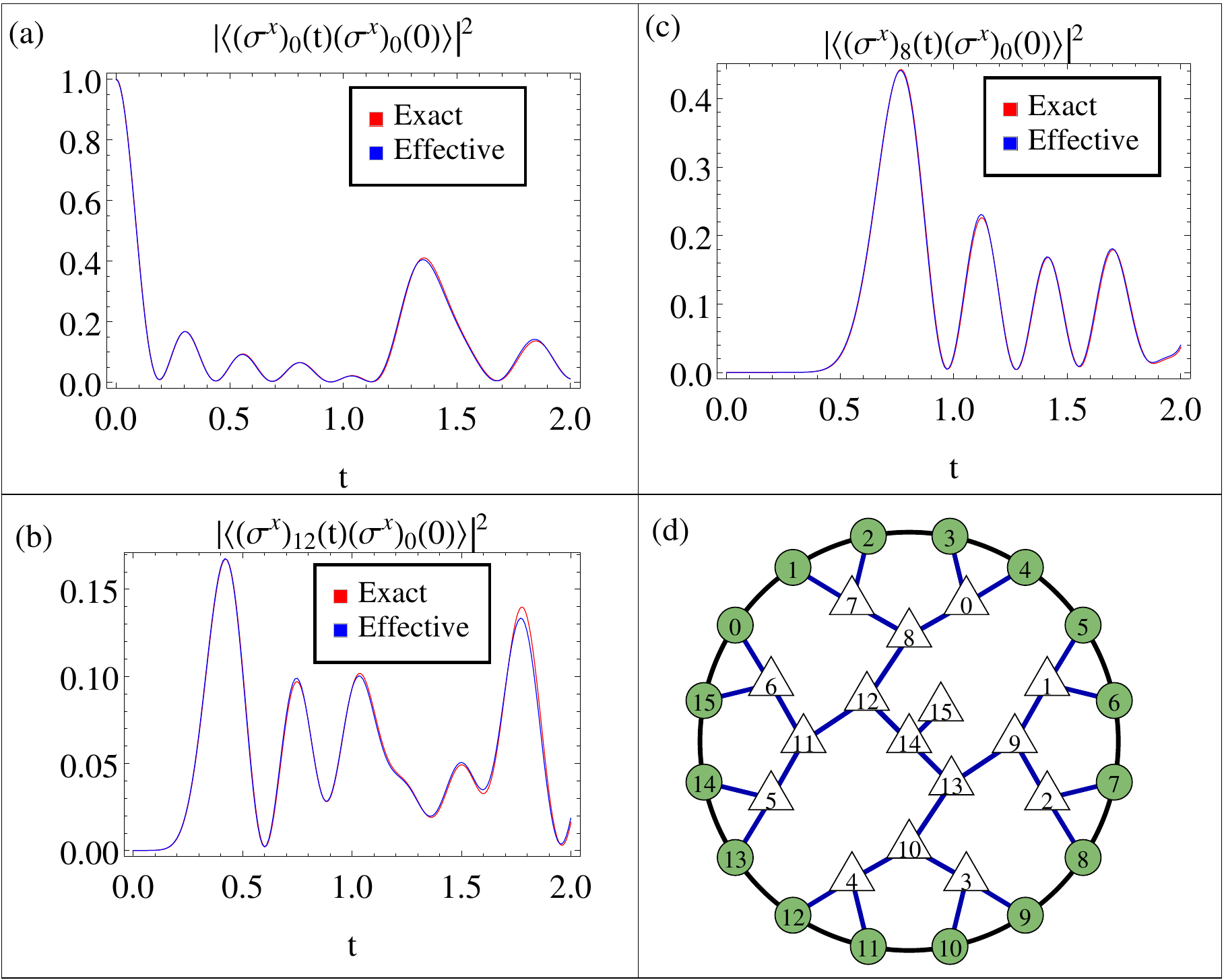}
	\caption{(color online) The square of the dynamical correlator $\langle \Omega_0|\sigma^x_{i}(\st)\sigma^x_{0}(0)|\Omega_0 \rangle$ computed exactly using the time-evolved many body wavefunction $|\Psi(\st)\rangle$ (red lines), and using the effective (blue line) 1-hologron Hamiltonian $H^{(1)}_\MM$ and the source coefficient $\{J_{i\mu}^x\}$ of Fig.\ref{fig:Jsources}. (a,b,c) Several time-dependent correlators from different physical sites of varying distance from $i=0$. (d) The MERA network with sites labeled for quick reference. }
	\label{fig:corr}
\end{figure}

Finally we relate the bulk correlator or propagator $\GG_{\mu\nu}(\st)$ and its physical or boundary counterpart $G_{ij}(\st)$ via (\ref{eqn:Gblk_Gbnd}) which we can compute explicitly using
\begin{align}
G_{ij}(\st) 
&= \langle 0| T \{J^x_i(\st) J^x_j(0)\} |0\rangle
\nonumber \\
&= \sum_n \sum_{\mu\nu} J^x_{i,\mu} J^x_{j,\nu} \langle \mu |\varepsilon_n \rangle \langle \varepsilon_n |\nu\rangle 
\e^{-i 2\pi(\varepsilon_n + E_1-E_0)|\st|}.
\end{align}
Shown in Fig.\ref{fig:corr} is the comparison between the dynamical correlation functions computed from just the physical degrees of freedom (using $|\Omega_0\rangle$ and $H$) and the holographic methods just described. Again there is excellent agreement between the exact and effective methods, demonstrating the usefulness of using the MERA network to compute time-dependent and out of equilibrium physical observables. Moreover, just like in Fig.\ref{fig:MRing}(e) sites further away from the excitation at $i=0$, require more time for correlations to build up from zero. Although this is naturally expected of the exact physical correlator, it is highly non-trivial that the bulk effective model produces this behavior too. However this is an important and necessary characteristic for any successful holographic duality.  

We should remark that although the computation of the correlator is made trivial by the fact that the transverse Ising model is exactly solvable via the JW transform, a correlated ground state is still invoked which is represented by a sum over an occupied Fermi sea of fermions. This is to be contrasted with the holographic approach where there is truly only a single body at play which is the excitation itself within the restricted subspace $\mathcal{H}^{(1)}_t$. 

\subsection{A Systematic Method for Constructing Effective Holographic Models}

\begin{figure*}
	\includegraphics[width=0.75\textwidth]{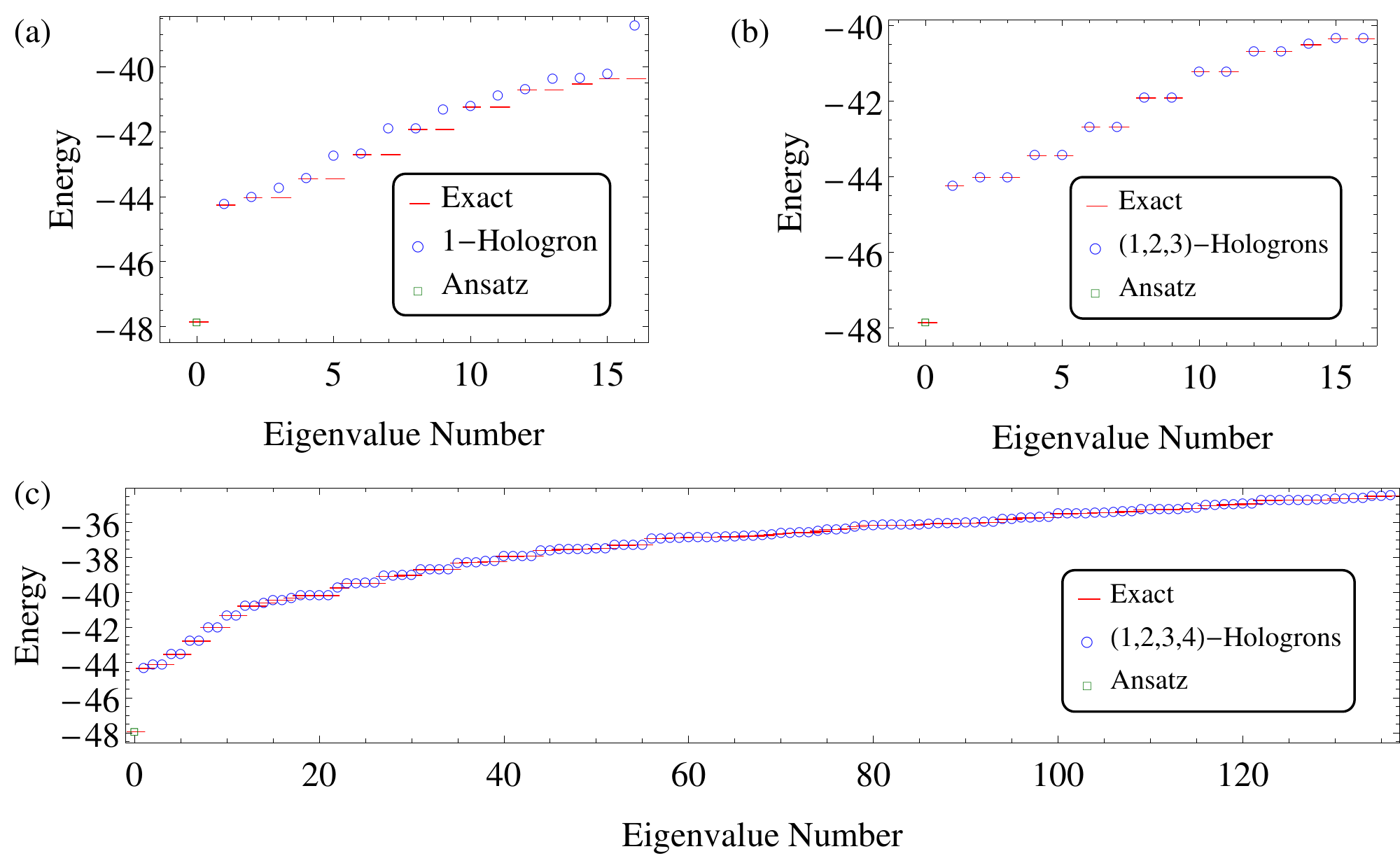}
	\caption{(color online) Comparisons between the exact physical energies and the effective holographic bulk Hamiltonian energies for a non-integrable deformation of the transverse Ising model. The model is generalized by adding a perturbation $V\sum_i \sigma^z_i \sigma^z_{i+1}$ to $H$. The specific parameters are $h^z=3$ and $V=0.1$ (a) The case where the bulk Hamiltonian contains at most a single hologron. (b) The case where up to three hologrons ($N_t \leq 3$) are included in the constructing the effective bulk Hamiltonian. The effective Hilbert space remains much smaller than $\text{dim}(\mathcal{H}_s)$, yet correctly captures the low energy spectra. (c) Comparison between the first 137 low energy exact eigenstates and the spectra computed with a hologron bulk Hamiltonian with up to four hologrons ($N_t \leq 4$).  The agreement or accuracy of the bulk effective hologron Hamiltonian greatly improves with increasing numbers of hologrons included.}
	\label{fig:energy2}
\end{figure*}

Now for the purposes of capturing the low energy effective physics using bulk degrees of freedom, the hologron basis provides a \emph{systematic} way to manufacture effective bulk Hamiltonians and improve their accuracy by enlarging the number of hologrons $N_t$ used. This is achieved by projecting $H_\MM$ into a fixed maximum number of $N_t$
\begin{align*}
\mathcal{H}^{\leq m}_t = \mathcal{H}^{(0)}_t 
\oplus \mathcal{H}^{(1)}_t \oplus \ldots \mathcal{H}^{(m)}_t \subset \mathcal{H}_t.
\end{align*}
Although unnecessary for exactly solvable models like the transverse field Ising model, such a construction may be invariably useful in non-integrable models. As a case in point, we consider a non-integrable version of the transverse field Ising model which includes the perturbation $H'=V\sum_i \sigma^z_i \sigma^z_{i+1}$. We then compare energy spectra obtained by exact diagonalization and from using an effective multi-hologron bulk Hamiltonian. This data is shown in Fig.\ref{fig:energy2}  and as expected, quantitative accuracy can be maintained from the effective approach whenever the number of degrees of freedom is sufficiently large. 

\subsection{Comparisons with the single mode approximation}

Finally, we should comment that the formulation of the low energy excited states using the hologron basis $\{|1_\mu\rangle \equiv \chi^x_{\mu}|0\rangle\}$ amounts to a single mode approximation (SMA) using the states $\{\sigma_i^x |\Omega_0\rangle\}$. However there are some key differences between the single mode approach and the MERA approach to building a low energy effective Hamiltonian. 

Firstly, the single excited modes $\{\sigma_i^x |\Omega_0\rangle\}$ do not form an orthogonal basis and a Gram-Schmidt orthogonalized basis which is not unique is needed. Moreover a Gram-Schmidt basis will in general lose the physical interpretation of a localized flipped spin, due to the linear superpositions involved. Often momentum-eigenstates from spin-flipped states are used as an orthogonal basis.  Furthermore systematically extending to a greater number of excited modes is less straightforward. By contrast the hologron basis, being a logical bulk degree of freedom, are always orthonormal and localized on the isometry tensors in the bulk. As was mentioned, the bulk hologron degrees of freedom are organized into different physical length scales and as with the case of a single hologron has a simple kinematic hopping Hamiltonian form in the bulk. In a sense, they are truly holographic.

Secondly, when computing the effective Hamiltonian in the single mode approximation, the following matrix elements have to be determined 
\begin{align*}
\langle \Omega_0 | \sigma^x_i H \sigma^x_j | \Omega_0\rangle, \quad \text{for all } i,j
\end{align*}
and is holographicaly dual to 
\begin{align*}
\langle 0| \chi^x_\mu H_\MM \chi^x_\nu | 0\rangle, \quad \text{for all } \mu,\nu.
\end{align*}
In the direct approach, the matrix elements of $H$ are simple but the ground state wavefunction $|\Omega_0\rangle$ is complicated and entangled. In the holographic approach, the bra and ket states are simple but $H_\MM$ is complicated and encodes the complexity that lies in $|\Omega_0\rangle$ through the tensors. This might not seem like a fair trade-off because of the added complexity of working with $H_\MM$. However in the cases where the ground state wavefunction $|\Omega_0\rangle$ is unknown or too large to store in memory, constructing low energy effective Hamiltonians from $H_\MM$ may be useful, provided that $\MM^\dagger|0\rangle$ can produce a good variational approximation to $|\Omega_0\rangle$.

\section{Emergent bulk geometries}\label{sec:emergent_geometry}

Finally we present some speculations about the emergent bulk background geometry which is probed by the 1-hologron sector. If we imagine taking the large system size limit, then we can expect that the 1-hologron dynamics as described by $h^{(1)}_\MM$ to approach a continuum Schr\"odinger operator. That is as $L \rightarrow \infty$ 
\begin{align}
h^{(1)}_{\MM} \rightarrow - \hat{\Delta} + \hat{V}
\label{eqn:h_continuum}
\end{align}
where ``lengths" are measured in appropriate units such that the kinetic term has the simplified Laplacian form $-\hat{\Delta}$. $\hat{V}$ is a local potential which we cannot rule out a priori in this context. One may regard that $-\hat{\Delta}$ contains only the kinematical information, while $\hat{V}$ the dynamical information. Strictly speaking, this conjecture requires that $h^{(1)}_\MM$ maintains its locality in the thermodynamic limit. Since our boundary theory here lies in its gapped phase which strongly satisfies the cluster decomposition principle, we expect that locality should extend to the bulk similarly. Although it remains a conjecture at this point that (\ref{eqn:h_continuum}) is a correct continuum limit. With this caveat, the geometry of the bulk can then be read off from $\hat{\Delta}$ in principle. This could perhaps be achieved by path-integral quantizing and determining the metric tensor from the kinetic part of the action. Finally, it has been discussed in the literature\cite{beny2013causal,czech2015integral,bao2015consistency}that the bulk geometry of a MERA network in the continuum limit, may actually correspond de-Sitter space-time instead of AdS. We expect that determination of $\hat{\Delta}$ and the geometry it describes will actually shed some light on this issue. 

\section{Conclusions}\label{sec:conclusions}

In this paper we have re-purposed the MERA tensor network as an analysis tool to study excited state evolution of the 1D transverse field Ising model. By reinterpreting MERA as a reversible quantum gate between physical and `logical' qubits, our numerical work has yielded bulk holographic dynamics that realizes in a concrete way, many of the features of the gauge-gravity (AdS/CFT) holographic duality; all without quantum gravity nor gauge symmetry. We have primarily studied the transverse field Ising model in its gapped paramagnetic phase ($h^z=3$) using the unitary transform provided by the MERA. We chose to do so because the strength of this correlated-uncorrelated duality improves the further away the transverse field Ising model is from its CFT point. Also the use of optimized disentanglers was key to realizing this duality; namely the fidelity between the exact ground state and the MERA ansatz. 

Our `holographic analysis' of a locally excited ground state and the first few excited eigenstates then yielded Hilbert space localization within the bulk Hilbert space. This then prompted the definition of a stable holographic bulk quasi-particle which we refer to as the hologron in this paper. These hologron excitations that reside in the ancillae qubits of the MERA quantum circuit, span the orthogonal complement to the MERA ansatz and are purely logical or informational in character. In the sense that they encode quantum fluctuations measured relative to the MERA ansatz which is a trivial product (vacuum) state in the bulk. Nevertheless, further study of hologrons in the single number sector, specifically their physical (boundary) characteristics, their sourcing by spin-flips $\sigma_i^x$, their relationship to the lowest manifold of excited eigenstates and their two dimensional bulk Hamiltonian and real-time dynamics have shown them to be interesting physical degrees of freedom living in one-dimension higher. More precisely, single hologrons were shown to be the holographic dual to the Ising `magnons' or spin-flips in the spin-chain. Then extending to the multi-hologron Hilbert space by truncating the maximum number of hologron in the bulk Hamiltonian produces ever more accurate effective low energy Hamiltonians. This provides a novel and systematic way to construct low energy effective Hamiltonians which was demonstrated to work even for a non-integrable deformation of the transverse field Ising model. 

It would be important to extend these results beyond the $L=16$ finite size. However, this will certainly require the use of MPS based techniques. 
We speculate that asymptotic forms for the MERA tensors and the low energy bulk Hamiltonian for arbitrarily long spin-chains may be derivable for large $h^z$ as an $1/h^z$ expansion.
Another interesting scenario would be to study in a controlled way, hologron-hologron scattering in longer chains. Other more obvious extensions would be to generalize to more complicated situations which perhaps exhibit non-trivial symmetry, topological order, finite temperature or models in higher dimensions. As for latter case, we predict that the low energy dynamics could also be captured using an adequately optimized MERA quantum circuit. Other possible directions for future work might include studying dynamics within the MERA circuit in the presence of symmetry. \cite{singh2010tensor} There have also been several recent studies \cite{yao2016discrete,von2016phase, zhang2016density, else2016floquet} involving stabilization of out-of-equilibrium or periodically driven Floquet systems in disordered systems. We expect signatures of such phenomena in the MERA bulk. Since our methods are suited to study dynamics, albeit in small systems, studying periodic drives would be a natural step forward.

\begin{acknowledgments}
We acknowledge the useful conversations with G Vidal. 
VC would like to specially thank H Changlani, V Dwivedi, S Swaminathan, S Ramamurthy, G Vanacore, A Narayan, N Tubman and S Young for useful and light-hearted discussions. AT would like thank R Leigh for a wonderful course on holography. VC very gratefully acknowledges the generous support from the Gordon and Betty Moore Foundation’s EPiQS Initiative through Grant GBMF4305 and the ICMT at UIUC. This work was supported in part by the National Science Foundation grant DMR-1455296, by the U.S. Department of Energy, Office of Science, Office of Advanced Scientific Computing Research, Scientific Discovery through Advanced Computing (SciDAC) program under Award Number FG02-12ER46875.

\end{acknowledgments}

\appendix

\section{MERA network construction.}\label{app:MethodDisIso}

\begin{figure}
	\boxed{\includegraphics[width=0.4\textwidth]{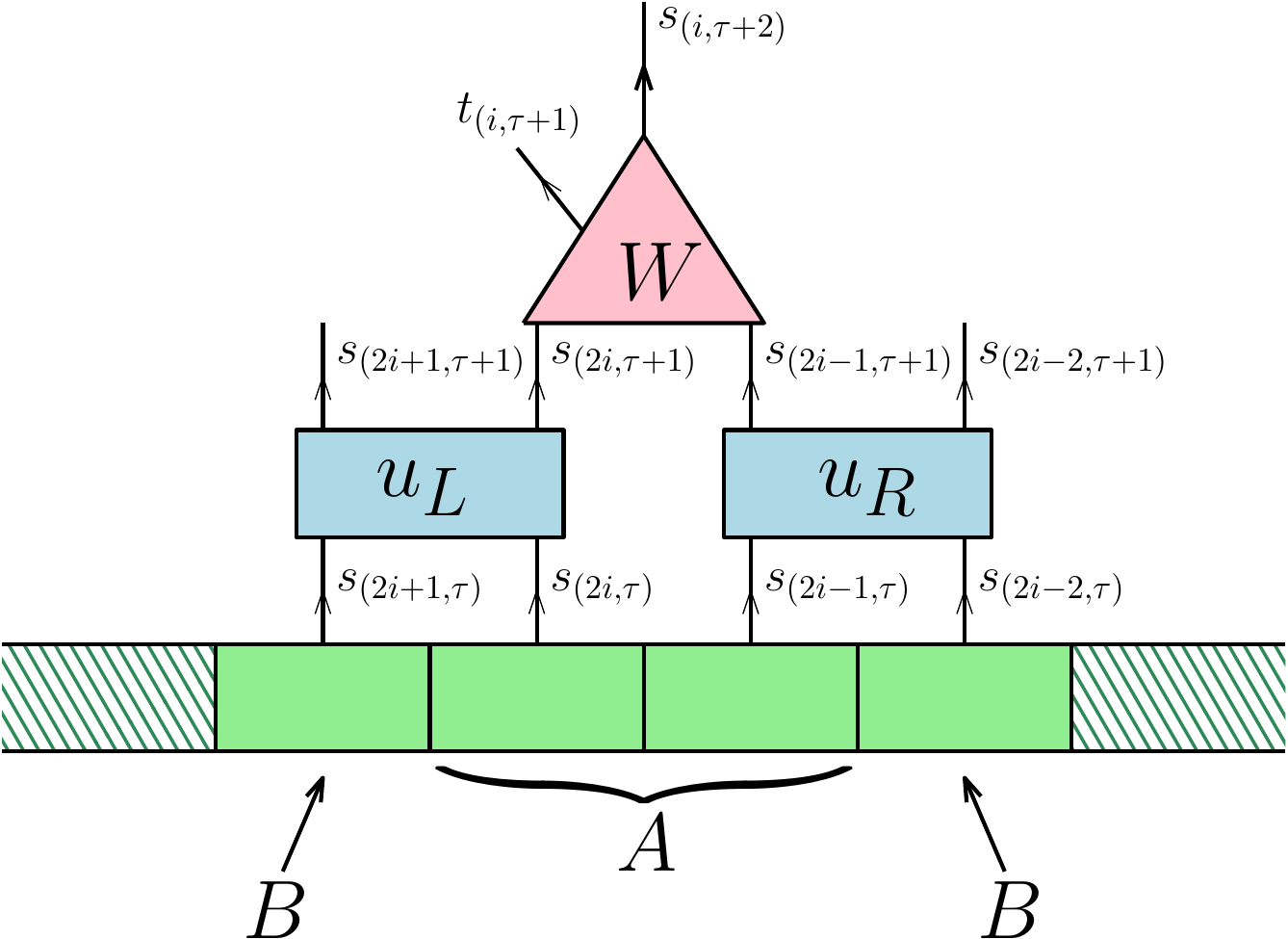}}
	\caption{ (color online) A basic 4 site block (solid green) acted by a left/right disentanglers $u_{L,R}$ and an isometry $W$. The hashed green regions denote the rest of the sites that act as the environment. Of the 4 sites, the central two sites denote the $A$ region, and the left/right sites, the $B$ region. A 4-site block reduced density matrix $\rho_0 ^{AB}$ is obtained by tracing over the environment (hashed) degrees of freedom. The left/right disentanglers act to minimize the short ranged entanglement between the central $A$ sites and its boundary $B$ sites. This entanglement is quantified by the von-Neumann entropy $S_\text{ent}[\rho^{A}_u]=-\text{Tr}(\rho_u^{A} \log \rho_u^A)$ of the disentangled $A$ reduced density matrix $\rho_u^{A} = \text{Tr}_B(u \rho_0^{AB} u)$ with $u=u_L \otimes u_R$. The $W$ isometry tensor then acts to separate the high and low entanglement Schmidt states producing new bulk $t_\mu$ and $s_\mu$ output qubits. Here $u_L \equiv U_{(i,\tau)}$, $u_R\equiv U_{(i-1,\tau)}$ and $W \equiv W_{(i,\tau+1)}$. } \label{fig:app_4block}
\end{figure}

In this appendix we describe the construction of the MERA quantum circuit from a ground state wavefunction previously obtained by exact diagonalization. The basic unit is a contiguous block of 4 sites obtained by partitioning the chain into units of 4, see Fig.\ref{fig:app_4block}. The physical state of this 4-site block state is given by a 4-site reduced density matrix (4RDM) $\rho^{AB}_0$  obtained by tracing out its environment. This 4RDM is acted on by a pair of left and right disentanglers $u :=u_L \otimes u_R \in \text{SU(4)} \times \text{SU(4)}$ giving $\rho_u^{AB} := u \rho^{AB}_0 u^\dagger$. Thereafter a disentangled 2-site reduced density matrix (2RDM) is obtained from tracing out the left and right most sites $\rho_u^A =\text{Tr}_B \rho^{AB}_u$. The isometry $W$ (unitary gate) is an SU(4) matrix that eigen-decomposes $\rho_{u}^A$ into new bulk ${t}_\mu$ and renormalized ${s}_\mu$ qubits, as described in Section \ref{sec:formulation}. These output states are organized by the eigenvalues of $\rho_u^A$ as shown in Fig.\ref{fig:app_rho}(c). This 4-site block process is iterated until no more 4-site blocks are available at that $\tau$ level. The total action of these gates for a single MERA level ($\tau$) produces the next MERA level ($\tau+1$). A new pure state wavefunction with half the number sites is obtained when the all the $\{{t}_{i,\tau}\}_i$ bulk qubits are projected onto their low $|0\rangle$ state. This entire process is iterated until the entire network of tensors is obtained.

The main challenge then is to compute the disentanglers $u_L, u_R \in $SU(4) which minimize the 2-site block entanglement entropy $S_\text{ent}[\rho_u^{A}]=-\text{Tr}(\rho_u^{A}\log \rho_u^{A})$. To achieve this goal we have employed a simple gradient descent minimization algorithm which we describe in the next subsection. The chief technical complication lies in computing the gradient of the $S_\text{ent}[\text{Tr}_B(u\cdot u^\dagger)]$ functional on 4RDMs, in the tangent space of SU(4)$\times$SU(4) unitaries. 

\subsection{The gradient of the $\mathscr{S}$ functional}

The total disentangler $u = u_{L} \otimes u_{R}$ is an element of SU(4)$\times$SU(4) $\subset$ SU(16) and $\rho_{0}^{AB}$ as a matrix is an element of $\mathbb{C}^{16 \times 16}$. The functional $\mathscr{S} : \text{SU(4)}\times\text{SU(4)} \rightarrow \mathbb{R}$ to be minimized is given explicitly by 
\begin{align}
\mathscr{S} (u_L,u_R) := S_\text{ent}\left[\text{Tr}_B\left(
(u_L \otimes u_R) \rho_0^{AB} (u_L\otimes u_R)^\dagger  \right) \right]
\end{align}
with
\begin{align}
S_\text{ent}[\rho] := -\text{Tr} (\rho \log \rho)
\end{align}
and where $\rho_0^{AB}$ is fixed. For numerical purposes we coordinatize SU(4) by its matrix elements in $\mathbb{C}^{4\times4}$, as $u_{L,R} = \sum_{I}(u_{L,R})_I E_I$ with $I=(i,j)$ as a multi-index and $E_I \equiv E_{(i,j)} := e_i \otimes e_j^T$. In these coordinates we have the Kronecker product
\begin{align} 
u&=(u_L)_I(u_R)_J E_I\otimes E_J \nonumber \\ 
&=: u_{IJ} E_{IJ} 
\end{align}
with the implied Einstein summation of repeated indices. 
Then taking unconstrained partial derivatives of $\mathscr{S}(u_L,u_R)$ with respect to $u_{L,R}$ and its complex conjugate $\overline{u}_{L,R}$ yields
\begin{align}
\frac{\partial \mathscr{S}}{\partial u_{IJ}} 
& = -\text{Tr}_{A} \left( \left[\mathbbm{1}+\log(\text{Tr}_{B}[u\rho_{0}^{AB}u^{\dagger}])\right]
\cdot \text{Tr}_{B}[E_{IJ}\rho_{0}^{AB}u^{\dagger} ] \right) \nonumber \\
& =  -\text{Tr}_{A}\left(F[u,\rho_{0}^{AB}]\cdot\text{Tr}_{B}[E_{IJ}\rho_{0}^{AB}u^{\dagger}]\right) 
\label{eqn:appDuS}\end{align}
and 
\begin{align}
\frac{\partial \mathscr{S}}{\partial \overline{u}_{IJ}} 
& = -\text{Tr}_{A} \left( \left[\mathbbm{1}+\log(\text{Tr}_{B}[u\rho_{0}^{AB}u^{\dagger}])\right]
\cdot \text{Tr}_{B}[u \rho_{0}^{AB}E_{IJ}^T ] \right) \nonumber \\
& =  -\text{Tr}_{A}\left(F[u,\rho_{0}^{AB}]\cdot\text{Tr}_{B}[u\rho_{0}^{AB}E_{IJ}^T]\right). 
\label{eqn:appDuS2}\end{align}
Out of convenience we have defined the auxiliary functional $F[u,\rho_{0}^{AB}] := \mathbbm{1}+ \log(\text{Tr}_{B}[u\rho_{0}^{AB}u^{\dagger}])$. Note that due to the $\log$, $F$ is only defined in the orthogonal complement to the kernel of $\text{Tr}_{B}[u\rho_{0}^{AB}u^{\dagger}]$. We conventionally take the functional trace-log to be zero on the kernel. 

The unitary constraints of SU(4)$\times$SU(4) require special consideration when attempting to minimize $\mathscr{S}$ via gradient descent. The method outlined by Manton\cite{Manton_disentangler_opt} addresses this problem in terms of optimization over the more general Stiefel manifolds. 
As a Stiefel manifold SU(4) $\subset \mathbb{C}^{4\times 4}$ is a nice compact and locally convex manifold inside the space of 4 $\times$ 4 complex matrices. Inside $\mathbb{C}^{4\times 4}$ one can define a projection $\Pi: \mathbb{C}^{4} \rightarrow \text{SU(4)}$ such that
\begin{align}
\Pi(x) = \text{arg} \underset{u \in U(4)}{\text{min}} ||x -u||_{HS}^{2}
\end{align}
with $||x -u||_{HS}^{2} = \text{Tr}[(x-u)^{\dagger}(x-u)]$ being the Frobenius or Hilbert-Schmidt norm.  
This projection map is provided by the Singular-Value Decomposition (SVD) of $x$ such that  if $x = U \Sigma V^{\dagger}$ then $\Pi(x) = UV^{\dagger}$ where $U,V \in $SU(4). The diagonal matrix $\Sigma$ has only non-negative entries. To see this consider 
\begin{align}
\text{Tr}[(x-u)^{\dagger}(x-u)] 
& =  \text{Tr}[(U \Sigma V^{\dagger} -u)^{\dagger}(U \Sigma V^{\dagger}-u)]  \nonumber \\
& =  \text{Tr}[V(\Sigma -U^{\dagger}uV)^\dagger U^{\dagger}U(\Sigma -U^{\dagger}uV)V^{\dagger}] \nonumber \\
& = \text{Tr}[(\Sigma -u')^{\dagger}(\Sigma -u')]
\end{align} 
where $u':= U^{\dagger}uV$ is unitary also. Since $\Sigma$ is diagonal and non-negative, the right hand side is minimized for $u'$ diagonal and real positive.  This is achieved only if $u' = \mathbbm{1}$ which implies $u = UV^{\dagger}$ as claimed.

Next we want to parametrize local neighborhoods of $x \in $SU(4) by the local tangent space.  Employing notions from Lie groups, we use the left invariant vector fields to describe the tangent space $T_{x}$SU(4).  We take $v:= (L_{g})_{*}v_{0} \in T_{x}$SU(4) where $v_{0}\in T_{e}$SU(4) and $L_{g}$ is the left translation (matrix multiplication) operator.  Concretely in $\mathbb{C}^{4\times 4}$ matrix coordinates if $x(0) = x$ and $\dot{x}(t) = v$, then $x(t) =x \exp(tv_{0})$ with $v = \dot{x}(0) = xv_{0}$.  Taylor expanding up $O(t^{2})$ yields
\begin{align}
x(t) & =  x(1 +tv_{o}) + O(t^{2}) \nonumber \\
& =  x +tv+ O(t^{2}) 
\end{align}
The special unitary constraint naturally requires the skew Hermiticity, $v_{0}^{\dagger} = -v_{0}$ and traceless-ness Tr $v_{0} =0$. The set of all such $v_{0} \in T_{e}$SU(4) naturally is the Lie algebra su(4) which is a $N^{2}-1 =15$ dimensional $\mathbb{R}$ vector space. Thus su(4) is our tangent space and the $\exp$ map provides a map to the local neighborhood 
\begin{align}
x\exp(\cdot) : \text{su(4)} &\rightarrow \text{SU(4)}
\nonumber \\
v_{0} &\mapsto x\exp(v_{0})
\end{align}
where $x \in$ SU(4). However instead of carrying out the full $\exp$ map numerically, we will use the projection $\Pi$ to map from the tangent space and local point SU(4)$\times$ su(4) to a nearby SU(4) point 
\begin{align}
(x,v) \mapsto \Pi(x(\mathbbm{1}+v)).
\label{eqn:appPi}\end{align}
We expect that locally $\Pi(x(\mathbbm{1}+t\,v_{0})) = x +t\,xv_{0} + \frac{t^{2}}{2}xv_{0}^{2} +O(t^{3})$ from just expanding the $\exp$ map. 
\newline 

In summary the general recipe is:
\begin{itemize}
	\item[(1)]{Compute from expressions (\ref{eqn:appDuS},\ref{eqn:appDuS2}) the steepest directions in $\mathbb{C}^{4\times 4}$ which lowers $\mathscr{S}$.}
	\item[(2)]{Project those directions onto su(4) by removing a trace and demanding skew-hermiticity.}
	\item[(3)]{Use the resulting su(4) tangent direction to map to a new point in SU(4) using equation (\ref{eqn:appPi}).}
\end{itemize}

In practice a refined form of the Newton's gradient descent is used to vary the step size. We have used the back-track line search algorithm\cite{press2007numerical} which is one simple implementation of this. Although other gradient based methods such as conjugate gradients may also be employed.

\subsubsection{Symmetry and Gauge Fixing}

Finally we should also mention the consequences of translational and internal gauge symmetry. Like MPS, there is always an internal redundancy or gauge symmetry in the non-uniqueness of the tensors. In our method, this arises from the details of steepest descent algorithm through the initial guess of disentanglers say. In being consistent, we initialize all our disentanglers from identity at the start of the steepest descent. Also because of translational and inversion symmetry of the physical model, we can strictly enforce the condition that all tensors within an RG layer $(\tau = 2n,2n+1)$ are identical. However, this still leaves a residual even-odd asymmetry because the choice of network topology explicitly breaks the translational symmetry of the physical chain down to a period two translation. See Fig.\ref{fig:Jsources}(a,b) for consequences of this. 

Nevertheless, it is perfectly reasonable to obtain other optimized MERA networks which are either not translationally symmetric or differ from ours by an internal gauge transform. We like to speculate that this is part of a discrete ``emergent diffeomorphsim invariance" of the bulk discrete ``gravity-like theory". 

\subsection{Quality of numerically determined MERA}

\begin{figure*}
	\includegraphics[scale=0.75]{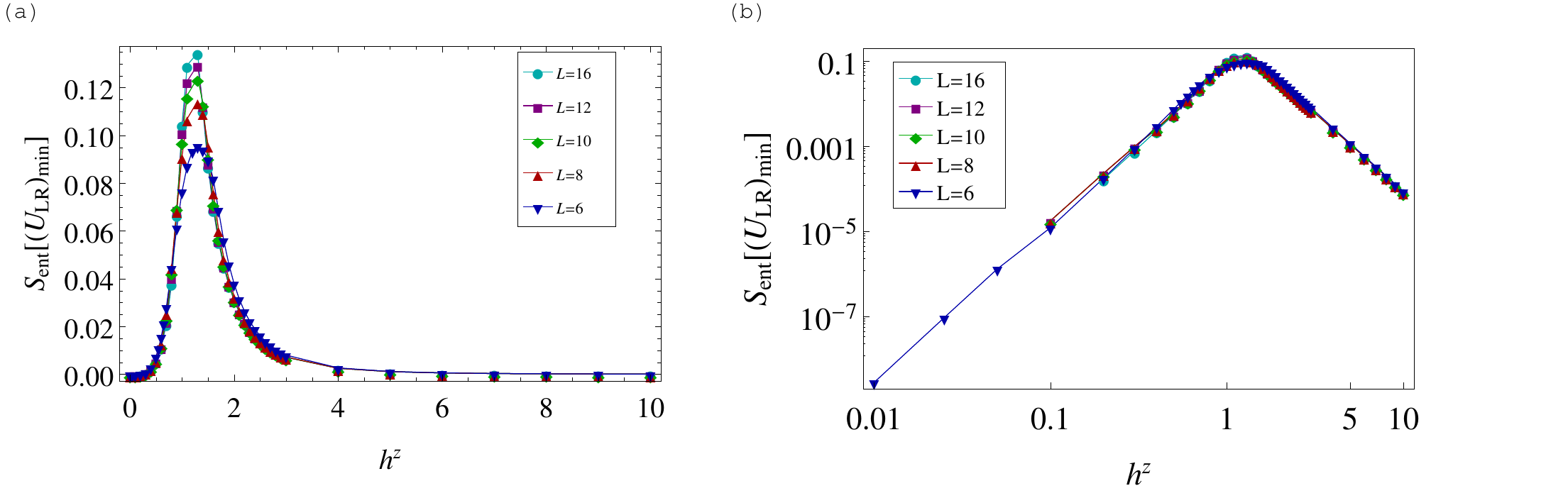}
	\caption{(a,b) The minimal entanglement entropy $S_\text{ent}$ of a 2RDM $\rho^{A}_u$ after optimal disentanglers have been applied at the first layer $\tau=1$. Data is shown in an (a) absolute  and (b) $\log$ scale. The 2-site block resides in the $\tau=1$ layer of the MERA and peaks near the critical point $h^z=1$, which is shifted due to finite size effects. Linear extrapolation of the maximum entanglement entropy from for log-log plot (b) yields $S_\text{ent}\approx 0.195$. This should be contrasted with the maximum possible entanglement entropy of $\log 0.25 \approx 1.39$ for 2 qubit density matrices.}
	\label{fig:Sent_2RDM}
\end{figure*}

We have used the method describe above to determine the isometry $\{W_\mu\}$ and disentangler $\{U_\mu\}$ tensors from numerical ground states. The quality of the resulting MERA tensors may be assessed by the amount of minimal residual 2-site block entanglement entropy that remains after the application of disentanglers. This residual entropy is greatest at the lowest or most fine-grained scale $\tau=1$ and continually decreases with increasing $\tau$ (RG iterations) in the MERA network. 

Shown in Fig. \ref{fig:Sent_2RDM} are computations of this minimal entropy at the $\tau=1$ level for various spin-chain lengths $L$ and transverse field $h^z$ values. As can be expected the optimized entanglement entropy is maximal near the critical point $h^z=1$, but the extrapolated maximum is shifted upwards from $1$ due to finite size effects. Nevertheless this entropy remains relatively small away from the critical point when deep in the gapped phases of the model, where it is expected that correlations are strongly classical ordered and the MERA network is efficient at removing short-range entanglement. Most of our numerical results use the value of $h^z=3$ deep in the paramagnetic phase where this residual entanglement entropy is small. Quantitatively, at the $\tau=1$ level without disentanglers the initial 2RDM block entropy is $8.55\times 10^{-2}$ and decreases to $7.14\times 10^{-3}$ with our optimized disentanglers, which represents a $92\%$ decrease. 

This is consistent with the expectation that classically ordered quantum states are ``close to" product states. The MERA circuit supports this by acting as an intermediary in the form of a finite depth quantum circuit that unitarily connects the numerical ground state to the product state $|0\rangle \in \mathcal{H}_t$ of $t_\mu$ qubits. 

\section{Quality of MERA at other points}

\begin{figure}
	\includegraphics[width=0.3\textwidth]{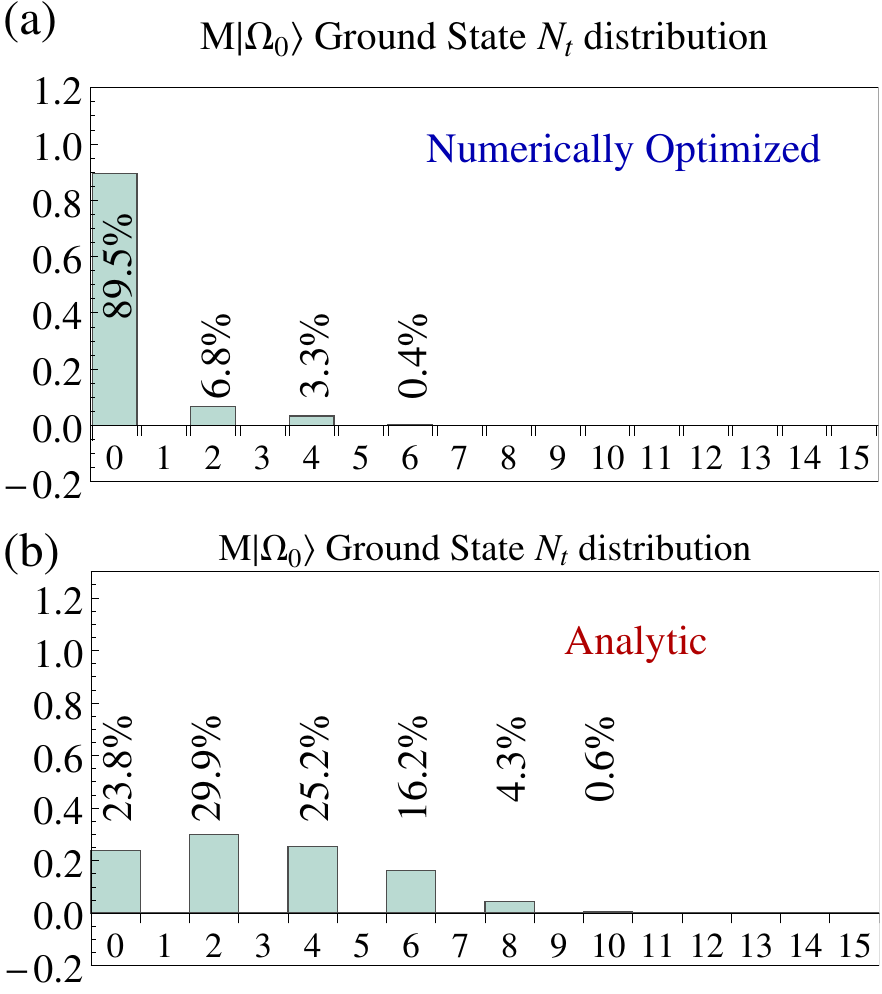}
	\caption{Spectral distribution of the total hologron number operator $N_t=\sum_\mu t_\mu$, for the transformed ground state $\MM|\Omega_0\rangle$ at the CFT point $h^z=1$ but at finite system size of $L=16$ with PBC. (a) The distribution using tensors of $\MM$ that are numerically optimized to maximize fidelity. (b) The distribution using the analytic expressions for the tensors from Evenbly and White\cite{evenbly2016entanglement}. }\label{fig:weights2}
\end{figure}

In this appendix, we show and comment on some results regarding our MERA tensor network $\MM$ for points other than the paramagnetic phase with $h^z=3$. The topology of the MERA network remains the same with $L=16$ and a constant bond dimension of $\chi=2$. The $\{W_\mu,U_\mu \}$ tensors were determined by the same numerical method described in Appendix \ref{app:MethodDisIso}. 

First, at the CFT point with $h^z=1$, a spectral decomposition with respect to the hologron number $N_t$ shown in Fig.\ref{fig:weights2}(a) reveals that the transformed ground state $\MM |\Omega_0\rangle$ is now significantly distributed over several hologron number sectors; namely the even $N_t$ subspaces. Note that the tensors have been re-optimized for the $h^z=1$ ground state. Thus in the details and for the system size of $L=16$, the MERA variational ansatz may differ in the high" energy  details. We however remark that finite-size effects in the $L=16$ system size and the constant bond dimension may be the limiting factor to optimizing the network. More recently Evenbly and White\cite{evenbly2016entanglement} have determined analytic expressions for the tensors of the scale-invariant (infinite size) limit of the transverse Ising model using the connection of MERA with wavelets. Shown in Fig.\ref{fig:weights2}(b) is the spectral distribution for the $h^z=1$ ground state using these expressions. They also show a large spread over hologron number space, but still limited to the even sectors. Again we attribute the low quality of the fidelity to the vacuum sector $|0\rangle$ due to finite-size effects, since these analytic expressions are known to produce the correct scaling dimensions of the Ising CFT.\cite{evenbly2016entanglement} Also the pattern of localization into only the even sectors is probably due a symmetry. 

\begin{figure}
	\includegraphics[width=0.3\textwidth]{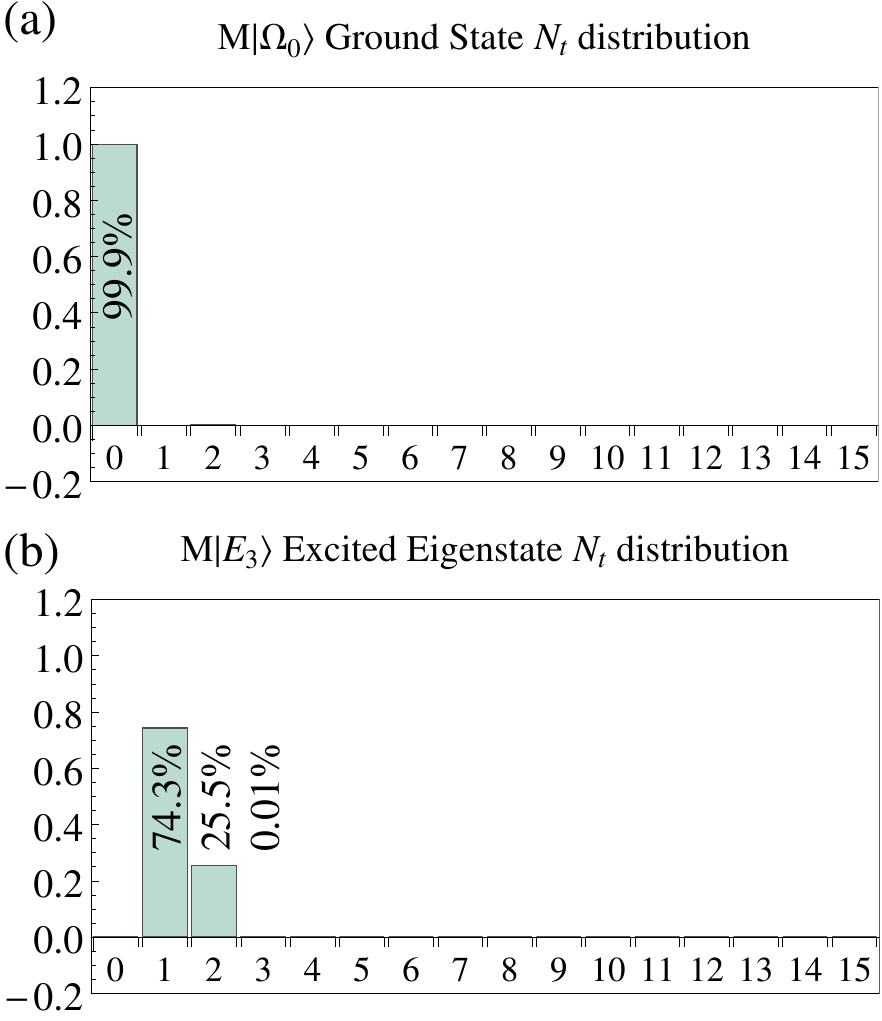}
	\caption{Spectral distribution of the total hologron number operator $N_t=\sum_\mu t_\mu$, for transformed eigenstates at the non-integrable point $(h^z,V)=(3,0.1)$. (a) The ground state's $\MM|\Omega_0\rangle$ spectral distribution is remains highly localized in the vacuum sector. (b) The third excited state spectral distribution is spread between $N_t=1$ and $N_t=2$ subspaces indicating a significant amount of hologron interactions which are not $N_t$ conserving. }
	\label{fig:weights3}
\end{figure}

Next, we also considered adding perturbations to the transverse Ising model, which breaks the exact integrability of the model. The specific perturbation considered is a $\sigma^z$$\sigma^z$ coupling defined by
\begin{align}
H' = V \sum_i \sigma^z_{i}\sigma^z_{i+1}.
\end{align}
The non-zero $V$ coupling effectively introduces weak interactions between Jordon-Wigner fermions of the 1D chain but we remain still in the paramagnetic phase. In this case the ground state $|\Omega_0\rangle$ remains well described by the ansatz $\MM^\dagger|0\rangle$ as shown in Fig.\ref{fig:weights3}(a) but the eigenstates are now spread beyond to neighboring number sectors. For example, in the third excited eigenstate $|E_3\rangle$ of $H+H'$, the spectral distribution now includes the $N_t=2$ sector as shown in Fig.\ref{fig:weights3}(b). This is indicative of hologron-hologron interactions in the bulk Hamiltonian $\MM (H+H')\MM^\dagger$ which are not $N_t$ conserving. As mentioned in the main text, computation of the effective bulk Hamiltonian and its energy spectra using up to the $N_t=3$ sector, reproduces well the exact spectrum in the low energy sector.

\section{Computing the bulk Hamiltonian.}\label{app:MethodHam}

In this appendix we briefly describe our numerical procedure to determine the matrix elements of the bulk Hamiltonian; which is not a common procedure. Now computing the matrix elements of $H_M = \MM H \MM^\dagger$ is a technical challenge, but it can be broken down to several smaller problems which amount to computing $L$ separate bond correlators with MERA tensors. Also it must be emphasized that although $H_M$ is much denser than $H$, it is much sparser than a dense $2^L \times 2^L$ Hermitian matrix. This is so because of the `causal cones' which bounds or kinematically constrains the communication between bulk qubits $t_\mu$. 

The first step is to express $H$ as a sum of bond terms between nearest neighbors, which for the transverse field Ising model is
\begin{align}
H  = \sum_{i=0}^{15} H_i, \quad \quad 
H_i = -\sigma^x_{i}\sigma^x_{i+1} + \frac{h^z}{2} \sigma^z_i.
\end{align}
Then for each $H_i$ we compute the matrix elements 
\begin{align}
\langle \nn | \MM \; H \;  \MM^\dagger |\nn'\rangle
\end{align}
for the hologron occupation states $|\nn\rangle, |\nn'\rangle$ which may span over a complete basis of $\mathcal{H}_s$ or a subspace such as the 1-hologron subspace $\mathcal{H}^{(1)}_t$. The influence of the MERA network appears in bra and ket states $(\langle \nn|\MM)$, $(\MM^\dagger| \nn '\rangle)$. Now from the discussion in Section \ref{sec:MERA_and_hologrons} regarding the physical meaning of the bulk qubit $t_\mu=0,1$ and hologrons in general, the state $\MM^\dagger|\nn\rangle$ is just the appropriate network with its isometries replaced with either low ($W_{l\mu}$) or high ($W_{h\mu}$) tensors depending on the configuration of $|\nn\rangle$. The resulting network is then contracted in the usual manner. In our computations we have relied heavily on the \href{http://itensor.org}{ITensor}\footnote{ITensor (\url{http://itensor.org}) is an easy to use but powerful tensor contraction library.} library to carry out these tensor network contractions which proceeds just like a bond energy calculation, but with the exception that the bra and ket states, or in the diagrammatic notation of Ref.\onlinecite {evenbly2009algorithms}, the top and bottom networks are not equal. Finally, the matrix elements for a fixed $H_i$, will only involve a fraction of the bulk sites. Namely those that lie in the bottom (past) and top (future) causal cones. This then leads to a sparser final $H_\MM$ than is naively expected. So for example, to compute the bond energy $\langle H_i\rangle $ of an arbitrary bulk wavefunction $|\psi\rangle$, one need only determine the reduced density matrix of $|\psi\rangle$ in the causal cone of $\MM H_i\MM^\dagger$, and then take a trace with the matrix elements of $\MM H_i\MM^\dagger$. 


\bibliography{MERA.bib}

\pagebreak

\widetext
\begin{center}
	\textbf{\large Supplemental Data}
\end{center}

\subsection*{Numerical Tensors}

\newcommand{\cw}{2.9cm} 
\newcommand\mc[1]{\multicolumn{2}{|c|}{#1}}

\begin{figure}[h]\centering
	\boxed{\includegraphics[width=0.14\textwidth]{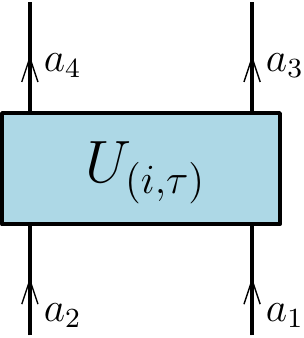} \quad \quad\includegraphics[width=0.15\textwidth]{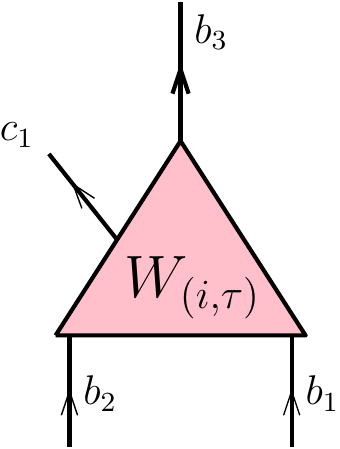}}
	\caption{Disentangler and isometry tensors with notation used for indices.}
	\label{fig:num_tensor_key}
\end{figure}

\noindent We tabulate here for reference purposes, our numerically obtained tensors from our variational procedure described in this section. These tensors were optimized for a chain of length $L=16$ and transverse field $h^z =3$ deep in the gapped paramagnetic phase. The notation used for the indices is described in Fig.\ref{fig:num_tensor_key}. In the isometry $W_\mu$ tables below, the first two rows correspond to the low isometry $W_{l\mu}$ and the last two row the high isometry $W_{h\mu}$. Due to translational symmetry, we only list a single isometry and disentangler per RG layer. 

\begin{table}[h!]
	\caption{$L_\tau=16$ layer}
	\begin{tabular}{|c|c||>{\centering}m{\cw}|>{\centering}m{\cw}|>{\centering}m{\cw}| c |}
		\cline{1-6}
		\mc{$U_{(i,\tau)}$} & \multicolumn{4}{|c|}{$a_2 a_1$}  \\ 
		\cline{3-6}
		\multicolumn{2}{|c|}{$\tau = 0$}
		&   00 & 01 & 10 & 11 \\
		\cline{1-6} 
		\multirowcell{4}{\begin{turn}{90} $a_4 a_3$ \end{turn}}
		&   00 & 0.99654154881 & 0 &  0 & 0.083095977579 \\
		&   01 & 0 & 0.99998132079 & 0.0061121251414 & 0 \\
		&   10 & 0 & -0.0061121251414  &  0.99998132079 & 0   \\
		&   11 &-0.083095977579 & 0 & 0  &  0.99654154881 \\
		\hline
	\end{tabular}
	\vspace{0.25cm} \\
	\begin{tabular}{|c|c||>{\centering}m{\cw}|>{\centering}m{\cw}|>{\centering}m{\cw}| c |}
		\cline{1-6}
		\mc{$W_{(i,\tau)}$} & \multicolumn{4}{|c|}{$b_2 b_1$}  \\ 
		\cline{3-6}
		\multicolumn{2}{|c|}{$\tau = 1$}
		&   00 & 01 & 10 & 11 \\
		\cline{1-6} 
		\multirowcell{4}{\begin{turn}{90} $c_1 b_3$ \end{turn}}
		& 00 &    0.99654810245 & 0 & 0 &    0.083017344619 \\
		& 01 &                0 &   -0.61517779795  &  -0.78838840485 & 0   \\
		& 10 &                0 &    0.78838840485  & -0.61517779795 &  0    \\
		& 11 &  -0.083017344619 & 0 &  0 &     0.99654810245 \\
		\hline
	\end{tabular}
\end{table}

\begin{table}[h!]
	\caption{$L_\tau=8$ layer}
	\begin{tabular}{|c|c||>{\centering}m{\cw}|>{\centering}m{\cw}|>{\centering}m{\cw}| c |}
		\cline{1-6}
		\mc{$U_{(i,\tau)}$} & \multicolumn{4}{|c|}{$a_2 a_1$}  \\ 
		\cline{3-6}
		\multicolumn{2}{|c|}{$\tau = 2$}
		&   00 & 01 & 10 & 11 \\
		\cline{1-6} 
		\multirowcell{4}{\begin{turn}{90} $a_4 a_3$ \end{turn}}
		& 00 &   0.99987281931 &               0  &               0  & 0.015948203953 \\
		& 01 &               0 &   0.99999894534  &  0.0014523480845 &               0 \\
		& 10 &               0 & -0.0014523480845 &  0.99999894534   &             0 \\
		& 11 & -0.015948203953 &               0  &              0   & 0.99987281931 \\
		\hline
	\end{tabular}
	\vspace{0.25cm} \\
	\begin{tabular}{|c|c||>{\centering}m{\cw}|>{\centering}m{\cw}|>{\centering}m{\cw}| c |}
		\cline{1-6}
		\mc{$W_{(i,\tau)}$} & \multicolumn{4}{|c|}{$b_2 b_1$}  \\ 
		\cline{3-6}
		\multicolumn{2}{|c|}{$\tau = 3$}
		&   00 & 01 & 10 & 11 \\
		\cline{1-6} 
		\multirowcell{4}{\begin{turn}{90} $c_1 b_3$ \end{turn}}
		& 00 & 0.99987263949 & 0 & 0  &  0.015959473428 \\
		& 01 & 0 & -0.61003633594 & -0.79237344026 & 0 \\
		& 10 & 0 & 0.79237344026 & -0.61003633594 & 0 \\
		& 11 & -0.015959473428 & 0 & 0 & 0.99987263949 \\
		\hline
		\end{tabular}
\end{table}

\begin{table}[h!]
	\caption{$L_\tau=4$ layer}
	\begin{tabular}{|c|c||>{\centering}m{\cw}|>{\centering}m{\cw}|>{\centering}m{\cw}| c |}
		\cline{1-6}
		\mc{$U_{(i,\tau)}$} & \multicolumn{4}{|c|}{$a_2 a_1$}  \\ 
		\cline{3-6}
		\multicolumn{2}{|c|}{$\tau = 4$}
		&   00 & 01 & 10 & 11 \\
		\cline{1-6} 
		\multirowcell{4}{\begin{turn}{90} $a_4 a_3$ \end{turn}}
		& 00 &   0.99999716508 &                0 &               0 &  0.0023811406335 \\
		& 01 &              0 &                1   &              0 &               0 \\
		& 10 &              0 &                0  &              1 &               0 \\
		& 11 & -0.0023811406335       &     0  &              0 &   0.99999716508 \\
		\hline
	\end{tabular}
	\vspace{0.25cm}\\
	\begin{tabular}{|c|c||>{\centering}m{\cw}|>{\centering}m{\cw}|>{\centering}m{\cw}| c |}
		\cline{1-6}
		\mc{$W_{(i,\tau)}$} & \multicolumn{4}{|c|}{$b_2 b_1$}  \\ 
		\cline{3-6}
		\multicolumn{2}{|c|}{$\tau = 5$}
		&   00 & 01 & 10 & 11 \\
		\cline{1-6} 
		\multirowcell{4}{\begin{turn}{90} $c_1 b_3$ \end{turn}}
		& 00 &    0.99999716514 & 0  & 0 &  0.0023811141786 \\
		& 01 &                0 & -0.70710678156 & -0.70710678081 & 0 \\
		& 10 &               0  &  -0.70710678081 &      0.70710678156 & 0 \\
		& 11 & -0.0023811141786 & 0 &  0 & 0.99999716514 \\
		\hline
		\end{tabular}
\end{table}	

\begin{table}[h!]
	\caption{$L_\tau=2$ layer. The disentangler is taken as the trivial identity matrix.}
	\begin{tabular}{|c|c||>{\centering}m{\cw}|>{\centering}m{\cw}|>{\centering}m{\cw}| c |}
		\cline{1-6}
		\mc{$W_{(i,\tau)}$} & \multicolumn{4}{|c|}{$b_2 b_1$}  \\ 
		\cline{3-6}
		\multicolumn{2}{|c|}{$\tau = 7$}
		&   00 & 01 & 10 & 11 \\
		\cline{1-6} 
		\multirowcell{4}{\begin{turn}{90} $c_1 b_3$ \end{turn}}
		& 00 &    0.99999998448 & 0  & 0 & 0.00017615432621 \\
		& 01 &               0 &   -0.70710678119 &     0.70710678119 & 0 \\
		& 10 &               0 &     0.70710678119 & 0.70710678119 & 0 \\
		& 11 & -0.00017615432621 & 0 &  0   &  0.99999998448 \\
		\hline
	\end{tabular}
\end{table}

\pagebreak

\subsection*{Animations}

In this supplemental section, we show several animated {\fontfamily{qcr}\selectfont .gif}'s to help the reader better visualize the bulk holographic MERA data. For playback, Adobe Acrobat$^{\tiny \textregistered}$ Reader 9.0 is recommended.

\subsection{Bulk hologron dynamics}\label{app:anime}

\begin{figure}[h!]
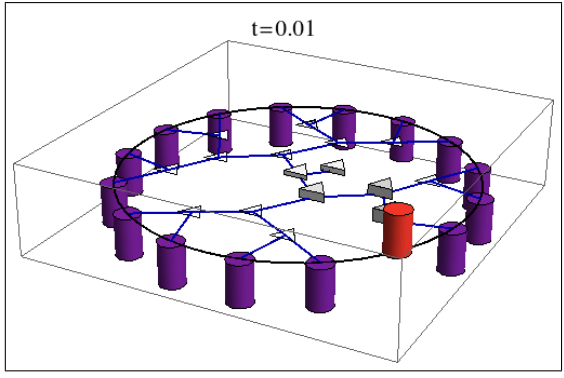

	\boxed{\animategraphics[width=0.4\textwidth,autoplay,loop]{12}{diagrams/MRing/L16_movie3D/MERA_L16_3D-}{0}{200}}
	\caption{ (colored animation) 3D representation of Fig.\ref{fig:MRing} which is the time development of a quenched ground state with $h^z=3$ for a periodic chain of length $L=16$. The system is excited by flipping a spin at position $i=8$ at time $t=0$. The height of the solid cylinders are proportional to $\langle \sigma_i^z\rangle$ which the boundary or physical spins. The heights of the solid triangles are proportional to $\langle t_\mu \rangle$ which are the bulk of hologron spins.  }
	\label{fig:MRing_anime}
\end{figure}

Here we present in Fig.\ref{fig:MRing_anime} an animation of the spin-flipped excited state analyzed under the MERA quantum circuit, for the data shown in Fig.\ref{fig:MRing}. 

\subsection{$|J_i^x\rangle$ Boundary insertions of 1-hologron states }\label{app:anime_J}

\begin{figure}[h!]
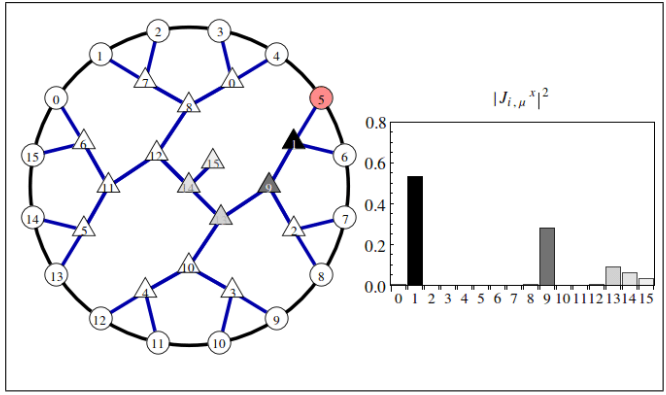

	\boxed{\animategraphics[width=0.475\textwidth,autoplay,loop]{1}{diagrams/Jx_movie/Jx-}{0}{15}}
	\caption{(color animation) Plots of the 1-hologron bulk states $|J_i^x\rangle$ sourced by the spin-flip operators $\sigma^x_i$ acting on ground state $|\Omega_0\rangle$. The red disk marks the spin-flipped site on the boundary and the histogram shows the probability distribution of the 1-hologron wavefunction in the bulk.}
	\label{fig:Jsources_anime}
\end{figure}

Fig.\ref{fig:Jsources_anime}  visualizes the 1-hologron states which are sourced by $\sigma^x_i$ excitations acting on the ground state. See section \ref{sec:sourcing}  and Fig.\ref{fig:Jsources}. 

\subsection{Physical States of Single Hologrons }\label{app:anime_Wave}

\begin{figure}[h!]
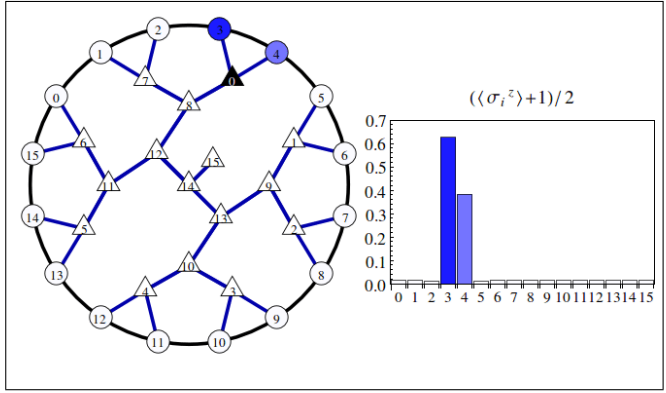

	\boxed{\animategraphics[width=0.475\textwidth,autoplay,loop]{1}{diagrams/Wave_movie/Wave-}{0}{15}}
	\caption{(color animation) Plots of the physical states of 1-hologron bulk states $\MM^\dagger \chi^x_\mu |0\rangle$. The black triangle marks the location $\mu$ of the hologron and the histogram shows the observable $(\langle \sigma^z_i\rangle +1)/2$ on the boundary.}
	\label{fig:Wave_anime}
\end{figure}

Fig.\ref{fig:Wave_anime} visualizes the physical state that local bulk hologron states produces. See Section \ref{sec:phys_hol}. 

\subsection{Bulk Eigenstates of the Single Hologron Excitations}\label{app:eigs}

\begin{figure}[h!]
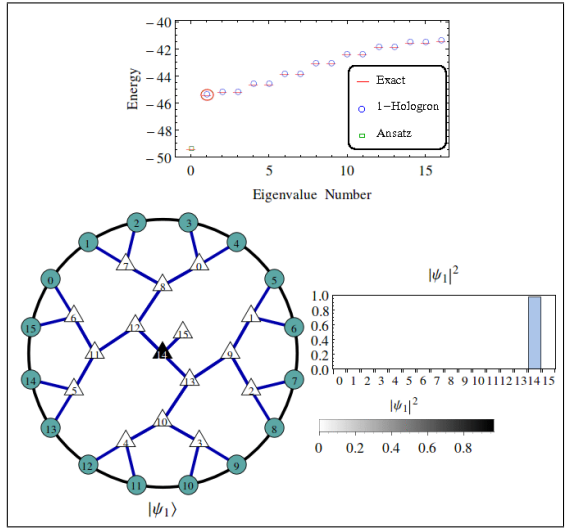

	\boxed{\animategraphics[width=0.4\textwidth,autoplay,loop]{1}{diagrams/Eig_movie/eigmovie-}{0}{15}}
	\caption{(color animation) Plots of the bulk 1-hologron of $H_\MM$ in the paramagnetic phase $h^z=3$.}
	\label{fig:eig_anime}
\end{figure}
Fig.\ref{fig:eig_anime} visualizes all the 1-hologron eigenstates as an animation. They demonstrate the general behavior that the higher energy eigenstates have greater weights among the UV bulk qubits $t_\mu$ nearer to the boundary, and vice versa. This again is consistent with the overall physical intuition that IR (UV) degrees of freedom are low (high) energy.

\subsection{Matrix Elements of $H_M^{(1)}$}

\begin{figure}[h!]
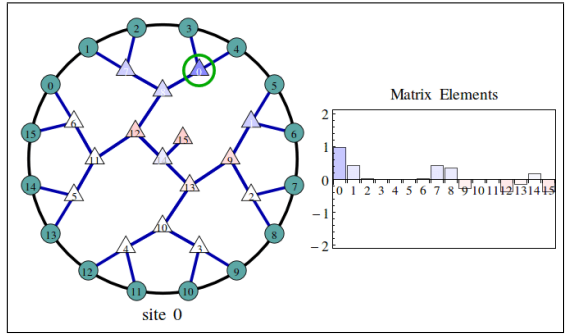

	\boxed{\animategraphics[width=0.4\textwidth,autoplay,loop]{1}{diagrams/ME_movie/ME-}{0}{15}}
	\caption{(color animation) Plots of the bulk 1-hologron of the matrix elements of $h_\MM$ in the integrable paramagnetic phase $h^z=3$.}
	\label{fig:ME_anime}
\end{figure}

In Fig.\ref{fig:ME_anime}, we present for completeness all the matrix elements of $h^{(1)}_M$ which is the traceless part of $H^{(1)}_M$ as interpreted as a hopping matrix element. The notation and symbolism follows that of Fig.\ref{fig:ME} in the main text.  

\subsection{Dynamical Correlators}

\begin{figure}[h!]
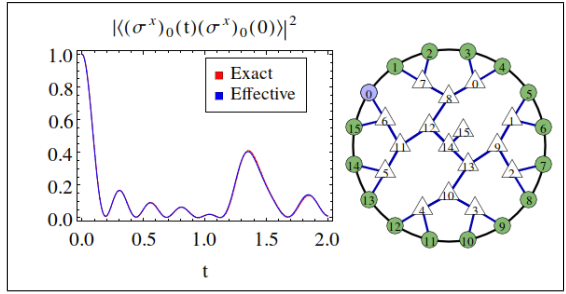

	\boxed{\animategraphics[width=0.4\textwidth,autoplay,loop]{1}{diagrams/Corr_movie/Corr-}{0}{15}}
	\caption{(color animation)  The square of the dynamical correlator $\langle \Omega_0|\sigma^x_{i}(\st)\sigma^x_{0}(0)|\Omega_0 \rangle$ computed exactly using the time-evolved many body wavefunction $|\Psi(\st)\rangle$ (red lines), and using the effective (blue line) 1-hologron Hamiltonian $H^{(1)}_\MM$ and the source coefficient $\{J_{i\mu}^x\}$ of Fig.\ref{fig:Jsources}.}
	\label{fig:Corr_anime}
\end{figure}

In Fig.\ref{fig:Corr_anime} we exhibit in an animation time-dependent $\langle \Omega_0 | \sigma^x_i(\st) \sigma^x_i(0)|\Omega_0\rangle$ correlator for all sites and their comparisons with numerically exact computation from the time-evolved physical wavefunction. 

\clearpage 

\end{document}